\documentclass[12pt]{iopart}
\usepackage{iopams}
\usepackage{graphicx}
\usepackage[T1]{fontenc}
\usepackage{ae}
\usepackage[latin1]{inputenc}
\usepackage{color}

\newcommand{\Eins}
           {\;\smash{\raisebox{-0.5ex}{$\!\!\stackrel{\!\mbox{1}
            \hspace{-0.4ex}\rule[0.0ex]{0.06ex}{1.60ex}}{ }$}}}
\newtheorem{lemma}{Lemma}

\newtheorem{ass}{Assumption}

\newtheorem{post}{Postulate}

\begin{document}


\title[GDQM]
      {Galilean decoherence and quantum measurement}
\author{Heinz-J\"urgen Schmidt$^1$}
\address{$^1$  Universit\"at Osnabr\"uck,
Fachbereich Mathematik, Informatik und Physik,
 D - 49069 Osnabr\"uck, Germany}


\begin{abstract}

In this study, we present a modified quantum theory, denoted as  $QT^\ast$,
which introduces mass-dependent decoherence effects.
These effects are derived by averaging the influence of a
proposed global quantum fluctuation in position and velocity.
While $QT^\ast$ is initially conceived as a conceptual framework - a ``toy theory" -
to demonstrate the internal consistency of specific perspectives in the measurement process debate,
it also exhibits physical features worthy of serious consideration.
The introduced decoherence effects create a distinction between micro- and macrosystems,
determined by a characteristic mass-dependent decoherence timescale, $\tau(m)$.
For macrosystems,  $QT^\ast$  can be approximated by classical statistical mechanics (CSM),
while for microsystems, the conventional quantum theory  $QT$ remains applicable.
The quantum measurement process is analyzed within the framework of $QT^\ast$,
where Galilean decoherence enables the transition from entangled states to proper mixtures.
This transition supports an ignorance-based interpretation of measurement outcomes,
aligning with the ensemble interpretation of quantum states.
To illustrate the theory's application, the Stern-Gerlach spin measurement is explored.
This example demonstrates that internal consistency can be achieved
despite the  challenges of modeling interactions with macroscopic detectors.

\end{abstract}

\maketitle
\section{Introduction}\label{sec:Intro}

The foundations of quantum theory, an extremely successful theory for 100 years,
continue to be the subject of controversy. A central aspect of this discussion
is the question of the ``semantic consistency" of quantum theory,
a term coined by C~.F.~von Weizs\"acker in \cite{W85}, chapter 11.2.
According to this author, ``the semantic consistency of a physical theory
should mean that its pre-understanding, with the help of which we physically interpret
its mathematical structure, itself satisfies the laws of the theory" (p.~514, my translation).
Applied to quantum theory, this raises the question of whether quantum mechanical
measurement processes themselves can in turn be correctly described using the tools of quantum mechanics.

We will try to roughly classify the different positions on this
``problem of the quantum mechanical measurement process".
The references to the position of individual physicists and their schools of thought
are intended for illustrative purposes only and not as a serious contribution to the history of physics.
One possible position, which we would like to call ``dualism",
would be the requirement to describe the measuring apparatus classically in principle,
whereby the cut between the measuring apparatus and the microsystem to be investigated can still be shifted.
This position could be associated with one of the variants of the so-called ``Copenhagen interpretation".
For the strict dualist, the measurement problem does not arise.
It can only be formulated if one attempts to analyze the measuring apparatus itself as a quantum mechanical system.

Historically, this attempt goes back to J.~von Neumann \cite{N32}.
A strong argument in favor of such an approach is atomism, i.~e.
the insight that macroscopic bodies (e.~g.~measuring apparatuses) are
made up of microsystems (e.~g.~atoms or molecules).
But even an atomist can doubt whether macroscopic systems can be described
by quantum theory in the same way as microsystems.
For example, in his early textbook \cite{L54}, G.~Ludwig discussed the problem
of the macroscopic arrow of time and relates this to the irreversible act of measurement.
One can imagine a more general theory that is responsible for micro- and macrosystems and contains
both the quantum mechanics of microsystems and, for example,
the classical mechanics of macrosystems as a limiting case, see for example \cite{L87}, \S  2.2.
This comprehensive theory should then also describe the quantum mechanical measurement
process in a semantically consistent way.
We will refer to the two non-dualistic positions that result from this as
``Q(uantum)-monism" and as ``G(eneralized)-monism".

For the Q-monist, the measurement problem presents itself in the following form:
After the interaction between the microsystem and the measuring apparatus has been completed,
the total system is in an entangled state, i.~e.~in a superposition of product states.
On the other hand, it appears that exactly one term of this superposition
is selected by reading the measurement result.
This transition from the superposition to a term is called ``reduction (or collapse) of the wave packet".
In a corresponding formulation of quantum theory,
there are therefore two different types of time evolution: 
the unitary, deterministic, continuous evolution based on the dynamics of the quantum system,
and the non-unitary, non-deterministic, discontinuous
evolution due to measurement and the associated collapse of the wave packet.
Because the latter cannot be reduced to the former,
quantum theory appears to be semantically inconsistent.

There are various attempts to solve the measurement problem within Q-monism.
One attempt, going back to von Neumannn \cite{N32}, London, Bauer and Langevin \cite{LB39},
combines the reduction of the wave packet with the act of consciousness
of reading the measurement result. Building on this, von Weizs\"acker
proposes to interpret the wave function $\psi$ as the quintessence
of the knowledge a physicist can have about a microsystem $S$.
Further measurements on $S$ then change this knowledge in a
comprehensible way, and von Weizs\"acker considers quantum theory
to be semantically consistent with this interpretation, see \cite{W85}.

Other approaches intended to solve the measurement problem
are the ``many worlds interpretation" of H.~Everett III  \cite{E57} and B.~DeWitt \cite{dW68},
Bohm's theory of hidden variables \cite{BB66},
the ``consistent history interpretation" \cite{G84},
the Ghirardi-Rimini-Weber theory of the collapse of the wave function \cite{GRW86} and
theories of decoherence \cite{Getal96}.
Furthermore, R.~Penrose's idea that the collapse of the wave packet
is triggered by gravitational effects \cite{P89}
should be mentioned here, although this is not elaborated further.

An alternative to these solutions would be to unmask the measurement problem as a pseudo-problem.
Although we are more concerned with the possible positions than with the people who hold them,
it should be mentioned here that this position can be found quite clearly
in the work of L.~E.~Ballentine \cite{B70} \cite{B15}.
To this end, he brings the interpretation of (pure or mixed)
quantum mechanical states into play. In the ``individual system interpretation",
a pure state provides a complete and exhaustive description of an individual system,
whereas in the ``ensemble interpretation" a (pure or mixed) state  provides
a description of certain statistical properties of an
ensemble of similarly prepared systems, \cite{B70}, p. 360.
Therefore, according to Ballentine, the paradoxical character of
the superposition of product states after the interaction
between microsystem $S$ and measuring apparatus $A$ disappears
in the ensemble interpretation. It no longer refers to a
fictitious state of the individual system $S+A$ but
to an ensemble of such individual systems.

It goes without saying that this is not the place to discuss the various proposed solutions in detail.
However, we would like to take a closer look at Ballentine's position outlined above.
To do so, we will, after some general remarks in \ref{sec:EIGR},
sketch the elaboration of the ensemble interpretation by Ludwig \cite{L85}
in subsection \ref{sec:EIPR}. Then we will argue, against Ballentine,
that the measurement problem also poses a difficulty in Q-monism with ensemble interpretation,
see subsection \ref{sec:EIAB}.

This leaves the supporters of the ensemble interpretation with G-monism.
However, one weakness of G-monism is that it is based on a theory that is
supposed to unite the micro and macro worlds, but which does not yet exist.
Is such a theory even possible?
What effects could play a role that only occur in a gradual transition from micro to macro systems?
Motivated by such questions, we propose in the main part of the paper,
starting with subsection \ref{sec:DEF}, as a toy theory
a modified quantum theory $QT^\ast$ that has the desired properties.
It uses elements from the decoherence theories, but differs from them
by assuming a global fluctuation in position and velocity (``Galilei fluctuation").
On states that are understood in terms of the statistical interpretation
these fluctuations act as averaged Galilean transformations
and thus lead to a mass-dependent decoherence, see the sections \ref{sec:PR} and \ref{sec:IDE}.
Similarly as in the well-known work on decoherence, this explains why certain
superpositions of states of heavy particles are not possible, see section \ref{sec:CW}
and \ref{sec:A1}.
The description of quantum measurement processes in the modified theory $QT^\ast$
is given in Section \ref{sec:QM}, but under a restrictive condition, see Assumption \ref{AM}.
In Section \ref{sec:ESG} we apply the theory $QT^\ast$ to the example of a
spin measurement by a Stern-Gerlach experiment. To ensure that the
interaction between the microsystem (spin) and the measuring apparatus
remains calculable, we simplify the detection of the silver atom by the
elastic collision with a macroscopic particle whose location can be
read off directly, see Section \ref{sec:SM}.
Despite this drastic simplification, the description of the
Stern-Gerlach experiment can be carried out consistently,
as shown by the choice of concrete numbers for the relevant
physical quantities in Section \ref{sec:CV} and \ref{sec:A2}.

Finally, our results are summarized and discussed in Section \ref{sec:S}.
In particular, we discuss whether the theory $QT^\ast$ outlined here
has physically plausible features beyond its toy character.

\section{Ensemble interpretation of quantum states}\label{sec:EI}
We will focus on the contributions to the ensemble interpretation of Ballentine \cite{B70} \cite{B15}
and Ludwig \cite{L85} \cite{L87}. For a broader overview see \cite{HW92}.

\subsection{General remarks}\label{sec:EIGR}

We begin with a few remarks that somewhat soften the supposed
contrast between ``individual system interpretation" and ``ensemble interpretation"
of quantum states. In doing so, we emphasize the differences to classical theories,
for which we consider classical statistical mechanics (CSM) as a generic example.
Thereby we ignore mathematical subtleties that are irrelevant to this discussion.
For example, pure states in CSM would be represented mathematically by points in phase space,
whereas in any physical application finite uncertainty arises.
Realistically, pure states are not realized by delta functions on
phase space but approximated by functions with a finite width.

First, we consider pure states in quantum theory that are represented by projectors
$P=|\psi\rangle\langle \psi|$ in a Hilbert space ${\mathcal H}$ with
$\psi\in{\mathcal H}$ and $\left\| \psi\right\| =1$.
If an ensemble of individual systems is prepared in a pure state $P$,
it is permitted to assign this pure state $P$ to each system $S$ of the ensemble in the sense
that any subsequent measurement on $S$ cannot lead to a contradiction to this assignment.
For example, a yes-no measurement with the projector $P$ will definitely return
the result ``yes" (or ``$1$") while another yes-no measurement with a projector
$Q\perp P$ would return the result ``no" (or ``$0$") with certainty.
All other yes-no measurements can result in ``$1$" or ``$0$"
without the possibility of predicting the outcome.
The difference to the CSM is that the existence of a pure state for an individual  system
is not only a permissible assumption here, but can also be fully verified by a single test measurement.
In quantum mechanics no single measurement can detect a pure state without using additional information,
see, e.~g., \cite{P93} \S\, 9.4.

Next, we consider genuinely mixed states, which are represented mathematically by statistical operators
$W$ in the Hilbert space ${\mathcal H}$, excluding the case $W=P$.
Here we have to distinguish between ``proper mixtures" and ``improper mixtures".
For proper mixtures, a partition of the corresponding ensemble into
sub-ensembles corresponding to pure states $P_i$, satisfying $W=\sum_i p_i\,P_i$,
can be distinguished due to the closer circumstances.
Again, it is allowed to assume that each individual system $S$ of the ensemble corresponding to $W$
``has" one of the states $P_i$, but one does not know which one (``ignorance interpretation").
This assumption cannot lead to a contradiction with the result of a subsequent measurement.
For example, in the case of mutually orthogonal $P_i$,
a measurement of an observable with the eigenprojectors $P_i$ will give the result
which of the pure states $P_i$ can be assigned to $S$, and is hence consistent with the assumption.
In CSM the situation is similar except that the partition of the ensemble into pure states
is unique and need not be given as an additional information for the test measurement.

In the case of improper mixtures, it is not possible to distinguish 
such a decomposition of the ensemble depending on the closer circumstances.
Below we will give an example where an assignment of pure states to individual systems of the ensemble can even be ruled out.
In CSM the case of improper mixtures does not occur since the decomposition of a mixed state into pure states
is always unique.

The announced example consists of the EPR-experiment, where a pair of particles is prepared in an entangled,
rotationally invariant spin state $\frac{1}{\sqrt{2}}(\uparrow \downarrow - \downarrow\uparrow)$
so that one particle is sent to Alice and the other one to Bob. If Alice makes a measurement
of the spin into direction ${\mathbf a}$, corresponding to the observable (Hermitean operator)
$A={\mathbf a}\cdot{\boldsymbol \sigma}$,
the probability of ``spin up" will be $1/2$ regardless of the direction ${\mathbf a}$.
Hence Alice will find that her particle has a mixed spin state described by $W=\frac{1}{2}\,\Eins$. The same applies to Bob.
Whether $W$ represents a proper or an improper mixture depends on the circumstances of the experiment.

If Bob has made a spin measurement of the observable $B={\mathbf b}\cdot{\boldsymbol \sigma}$
then it would be legitimate to call the mixed state $W$ of the particle sent to Alice ``proper"
and the corresponding decomposition $W=\sum_i p_i\,P_i$ is given by the two eigenprojectors $P_\uparrow, P_\downarrow$
of $B$, with $p_\uparrow=p_\downarrow=1/2$. According to the above it would be  permissible to
assume that each particle arriving at Alice ``has" either spin up or spin down w.~r.~t.~the direction ${\mathbf b}$
chosen by Bob and corresponding to the perfect anti-correlation of spin measurement results of Alice and Bob.

The situation is different in the case where Bob has not made any spin measurement. Although the particles
sent to Alice are characterized by the same mixed state $W=\frac{1}{2}\,\Eins$ as before, this mixture has now
to be classified as ``improper".
Imagine that Alice adopts a ``double ignorance interpretation" in the sense that she assumes a
proper mixture of the form $W_A=p_\uparrow P_\uparrow + p_\downarrow P_\downarrow$,
but without knowing the direction of $\uparrow$. The analogous assumption will be made by Bob.
To account for the anti-correlations as much as possible, Bob should choose the same unknown spin direction as Alice.
Both assumptions can be combined so that, taking into account the anti-correlation, the overall state has the form
$\widetilde{W}=\frac{1}{2} P_{\uparrow\downarrow} +\frac{1}{2} P_{\downarrow\uparrow}$.
This results in a mixture of product states for the combined system,
which is different from the pure state given by the projector $V$ on
$\frac{1}{\sqrt{2}}(\uparrow \downarrow - \downarrow\uparrow)$.
However, the difference between $V$ and $\widetilde{W}$ with unknown direction of $\uparrow$
cannot be detected by a single measurement, but only by a long series of measurements.

We conclude that the ``individual system interpretation" of quantum states is possible in some cases,
namely for pure states and proper mixtures, but for the general case, including improper mixtures, the
``ensemble interpretation" is indispensable. In this respect, our argument goes beyond
Ballentine \cite{B70}, who sees in the ``individual system interpretation"
a conceivable but superfluous supplement to the ``ensemble interpretation",
against which, among other things, Ockham's razor argument speaks.

\subsection{Preparation and registration procedures}\label{sec:EIPR}

We have seen in the preceding subsection that the distinction between ``proper" and ``improper"
mixtures cannot be made at the level of statistical operators $W$ but depends on the ``closer circumstances" of the experiments.
This can be taken as a motivation for introducing a refined vocabulary: A state described by $W$ is not
represented by a single ensemble of identically prepared systems but by a class of ensembles corresponding to
different ``preparation procedures". Such a refined theory has been elaborated by Ludwig \cite{L85} and
need not be reproduced here in all details. We will focus on the aspects relevant to this paper.

The attempt to interpret a quantum mechanical state as a single ensemble $E$ of
quantum systems leads to another problem:
Every conceivable measurement of an observable (Hermitian operator) $A$,
which has a complete system of eigenprojectors, leads to a corresponding partition of $E$
consisting of subsets of $E$ where the observable $A$ has a definite value.
For arbitrary finite sets of non-commuting observables
and a dimension of the Hilbert space $\ge 3$, however,
the existence of these partitions of a single ensemble $E$ leads to a
contradiction according to the theorem of Kochen and Specker \cite{KS67}.

A characteristic feature of Ludwig's approach is the extension to preparation and registration procedures.
Both types of procedures contain the description or construction manual of macroscopic devices
in a classical language and can be realized by concrete processes as often as required.
A single experiment consists of a combination of a realization of a preparation procedure
and of a registration procedure and can be in turn be repeated as often as you want.
The problem arises that there are obviously different combinations
that differ in the relative spatio-temporal position of their parts.
Ludwig solves this problem by demanding that the construction manuals refer to an abstract
coordinate frame, which is replaced by a concrete coordinate frame for each realization.
If you combine a preparation procedure and a registration procedure, you only need to identify the abstract coordinate frames.
Another advantage of the introduction of abstract coordinate frames is that it
enables a natural definition of time-translations or, more general Galilei transformations,
operating on the coordinate frames of preparation procedures (Schr\"odinger picture) or, alternatively, of registration procedures
(Heisenberg picture). 

The purpose of an experiment is to obtain results. Ludwig's formalism takes this aspect into account
by assigning a probability $\lambda(a,b_0,b)$ to the triple of a preparation procedure $a$,
a registration procedure $b_0$ and an event $b$ that can occur during the measurement.
This leads to a function
$\mu: {\mathcal Q}\times {\mathcal F}\rightarrow [0,1]$,
where ${\mathcal Q}$ is the set of (non-empty) preparation procedures and
${\mathcal F}$ the set of ``effect processes" consisting of pairs $(b_0,b)$ and
describing generalized yes-no measurements. Different events $b_i$ for the same registration procedure $b$
give rise to ``coexistent effect processes".
Each preparation procedure therefore induces a real function defined on ${\mathcal F}$, but two different
preparation procedures can induce the same function. Proceeding to the corresponding equivalence classes of
preparation  procedures, one obtains ``states" (we have avoided Ludwig's notion of ``ensembles"
as it could be confused with the previously discussed concept). Analogously, one considers
``effects" as equivalence classes of statistically equivalent effect processes,
and a derived probability function
\begin{equation}\label{KLmu}
 \mu: K\times L \rightarrow [0,1]
 \;,
\end{equation}
where $K$ denotes the set of states and $L$ the set of effects.
Coexistent effect processes generate ``coexistent effects".
In his work \cite{L85} Ludwig further provides physical axioms for the structure
$(K,L,\mu)$ such that $K$ can be represented as
the set of statistical operators $W$ on a Hilbert space ${\mathcal H}$ and $L$ as the set of
Hermitean operators $F$ on ${\mathcal H}$ such that $0\le F \le \Eins$ and the statistical
function $\mu$ is represented as $\mbox{Tr} (W F)$.

The description of a particular experiment with the terms developed by Ludwig can sometimes be done
in different ways, which is reminiscent of the shiftability of the cut in the Copenhagen interpretation.
Consider, for example, the EPR experiment mentioned in the section \ref{sec:EIGR}
of two measurements of Alice and Bob, resp.~, on a pair of particles.
On the one hand, these measurements can be understood as coexistent effects measured for a
pure state given by the projector onto $\frac{1}{\sqrt{2}}(\uparrow \downarrow - \downarrow\uparrow)$.
On the other hand, the preparation of the entangled pair together with the measurement of Bob
can be understood as the preparation $a$ of a mixed state $W=\frac{1}{2}\,\Eins$
on which Alice performs a measurement.
The decomposition of the set of individual systems prepared according to $a$,
corresponding to the result ``yes" or ``no" of Bob's measurement,
can be understood as a decomposition of $a$ into two partial preparation procedures,
$a=a_1 \cup a_2$,
which further leads to the convex linear combination
$W=\frac{1}{2} P_\uparrow+\frac{1}{2} P_\downarrow$ at the level of the states.
We have referred to this as a ``proper mixture" in subsection \ref{sec:EIGR}.

\subsection{Ensemble interpretation and the measurement problem}\label{sec:EIAB}

To fix the notation we will assume that, after the interaction between the microsystem and the measurement
apparatus, the total system assumes the pure state
\begin{equation}\label{stateafterinteraction}
 \Phi=\sum_{\alpha=1}^r c_\alpha |\Psi_\alpha\rangle \otimes |\psi_\alpha\rangle
 \;,
\end{equation}
where the states $|\Psi_\alpha\rangle$ refer to the microsystem and the states $|\psi_\alpha\rangle$
to the various outcomes which can be read off the measurement apparatus. Ballentine has argued \cite {B70} that this only then
contradicts the experience that the measuring apparatus shows an unambiguous result
if one assumes the individual system interpretation for the pure state $\Phi$.
There is supposedly no such problem with the ensemble interpretation for $\Phi$.

We will discuss this position using Ludwig's elaboration of the ensemble
interpretation outlined in the last subsection. Then the state $\Phi$ is
realized by a preparation procedure $a$.
One could of course object that the preparation of a pure initial state for the
measuring apparatus, which leads to (\ref{stateafterinteraction}) after time $t$, is extremely unrealistic.
But this would be an argument against Q-monism.
The discussion we analyze here, however, takes place within the framework of Q-monism,
and is thus based on the premise that such a preparation procedure $a$ is conceivable.

In Ludwig's language, the states $\psi_\alpha$ of the measuring apparatus
can be interpreted as mutually exclusive coexistent effect processes,
which enable a disjoint decomposition of the preparation procedure $a$ of the form $a=\bigcup_\alpha a_\alpha$,
similar to the EPR example from the subsection \ref{sec:EIPR}. 
But this would result in a mixed state $\Phi$,
in contradiction to the assumption that $\Phi$ is a pure state.

We therefore come to the conclusion that the measurement problem persists,
regardless of the interpretation of the state $\Phi$
according to the individual system interpretation or the ensemble interpetation.

\section{Modified Quantum Theory QT*}\label{sec:QT}

\subsection{Definitions}\label{sec:DEF}

We consider non-relativistic quantum theory based on Galilean space-time ${\mathcal M}$
which is assumed to be a $4$-dimensional affine space. Let ${\sf G}$ denote the (proper) Galilei group
acting on ${\mathcal M}$ by means of affine bijections $g:{\mathcal M}\rightarrow {\mathcal M},\,g\in{\sf G}$.
W.~r.~t.~a fixed Galilean coordinate system
$\gamma:{\mathcal M}\rightarrow {\mathbb R}\times {\mathbb R}^3,\,\gamma(m)=(t,{\mathbf x})$,
Galilean transformations can be identified with the quadruple
$g=(b,{\mathbf a},{\mathbf u},R)\in{\mathbb R}\times {\mathbb R}^3 \times {\mathbb R}^3\times SO_3$
such that
\begin{equation}\label{Galtrafo}
 g(t,{\mathbf x}) = \left(t+b, R\,{\mathbf x}+ {\mathbf u}\,t +{\mathbf a}\right)
 \;.
\end{equation}

The connection to quantum theory is given by a projective,  unitary, irreducible  representation (``irrep") ${\sf R}$ of ${\sf G}$.
This representation is characterized by a positive real number $m$ (``mass"), a positive half-integer number $s$ (``spin"),
and a Hilbert space
\begin{equation}\label{Hilbert1}
 {\mathcal H}_{m,s}={\mathbb C}^{2s+1} \otimes L^2\left({\mathbb R}^3, d^3{\mathbf p}\right)
 \;,
\end{equation}
such that
\begin{eqnarray}\nonumber
 && \left({\sf R}(b,{\mathbf a},{\mathbf u},R)\,\psi \right)_\alpha ({\mathbf p})\\
 \label{irrepGal}
 &&=
  \exp\left(\frac{{\sf i}}{\hbar}(Eb-{\mathbf p}\cdot{\mathbf a})  \right)
  \sum_{\beta=-s}^{s}D_{\alpha\beta}^s(R)\,\psi_\beta\left(R^{-1}({\mathbf p}-m{\mathbf u}) \right)
  \;,
\end{eqnarray}
see, for example, \cite{L76}.
Here $E=\frac{{\mathbf p}^2}{2m}$ and $D^s$ is the $(2s+1)$-dimensional projective,  unitary irrep
of $SO_3$ or, equivalently, the  $(2s+1)$-dimensional  unitary irrep of $SU_2$.

For the following we neglect the rotational degrees of freedom and thus set $s=0$.
Time translations are not considered since they correspond to a free time evolution and we will instead
consider a more general unitary, Galilei invariant time evolution $U(t)$.
Moreover, for the purposes of this paper it will be convenient to pass from the momentum representation
to the velocity representation of wave functions and hence to consider the Hilbert space
\begin{equation}\label{Hilbert2}
 \widehat{\mathcal H}_{m}=L^2\left({\mathbb R}^3, d^3{\mathbf v}\right)
 \;,
\end{equation}
Then (\ref{irrepGal}) will be reduced to
\begin{equation} \label{irrepGalredv}
 \left({\sf R}(0,{\mathbf a},{\mathbf u},{\Eins})\,\psi \right) ({\mathbf v})=
  \exp\left(-\frac{{\sf i}}{\hbar}m{\mathbf v}\cdot{\mathbf a}  \right)
  \psi\left({\mathbf v}-{\mathbf u} \right),\quad \psi\in  \widehat{\mathcal H}_{m}
  \;.
\end{equation}
In the position representation of wave functions obtained by an inverse Fourier transformation
we use the Hilbert space
\begin{equation}\label{Hilbert3}
{\mathcal H}_{m}=L^2\left({\mathbb R}^3, d^3{\mathbf x}\right)
 \;.
\end{equation}
Explicitly, the transformation from $\phi\in{\mathcal H}_{m}$ to $\psi\in\widehat{\mathcal H}_{m}$
can be written as
\begin{equation}\label{postovel}
  \psi({\mathbf v})=\left(\frac{m}{2\pi\hbar}\right)^{3/2}
  \int  \exp\left( -\frac{\sf i}{\hbar}m {\mathbf v}\cdot{\mathbf x}\right)\,\phi({\mathbf x})\,d^3{\mathbf x}
  \;.
\end{equation}

Then we obtain the positional representation of the reduced Galilean irrep analogous to
(\ref{irrepGalredv}) (denoted by the same letter)
\begin{equation} \label{irrepGalredx}
 \left({\sf R}(0,{\mathbf a},{\mathbf u},{\Eins})\,\phi \right) ({\mathbf x})=
  \exp\left(\frac{{\sf i}}{\hbar}m{\mathbf u}\cdot{\mathbf x}  \right)
  \phi\left({\mathbf x}-{\mathbf a} \right)),\quad \phi\in  {\mathcal H}_{m}
  \;.
\end{equation}

\subsection{Postulates and first results}\label{sec:PR}

So far, we have only sketched the well-known form of non-relativistic quantum theory.
We will modify this theory by postulating that the ``pure" time evolution described by a
unitary family $U(t)$ that commutes with the reduced Galilean irrep (\ref{irrepGalredx})
is superimposed by a global quantum fluctuation:
\begin{post}\label{P1}(Preliminary form)\\
Between the points in time $t$ and $t+\delta t$, the total time evolution,
in addition to the pure time evolution  $U(\delta t)$, is given by random, independent spatial translations ${\mathbf a}$
and Galilean boosts ${\mathbf u}$. $\delta t>0$ is a fixed time difference
that is very small compared to the typical duration $\Delta t$ of measurements.
\end{post}

The distribution of the two random variables  ${\mathbf a}$ and ${\mathbf u}$ is left open,
except that it is identical for all intervals $(t, t+\delta t)$ and will be later
restricted by certain assumptions on its statistical parameters. The order of space translations and boosts
is irrelevant since, according to Weyl's commutation relation,
the commutator of their irreps is a constant phase factor that cancels if calculating
probabilities.

We will discuss the consequences of Postulate \ref{P1} for the statistical interpretation of quantum theory,
as outlined in subsection \ref{sec:EIPR}.
We will concentrate on the influence of random translations and boosts on the ensemble $W$, since the pure
time evolution commutes with  their representation. For two different time intervals $(t_1,t_1+\delta t)$ and $(t_2,t_2+\delta t)$
the values of the random variables  ${\mathbf a}$ and  ${\mathbf u}$ and their irreps ${\sf R}(0,{\mathbf a},{\mathbf u},{\Eins})$ will be different.
But due to the statistical interpretation of the state $W$ the individual Galilean fluctuations are not important but
only the mean value of the corresponding irreps over many time intervals $(t,t+\delta t)$.
Additionally we use the assumption $\delta t << \Delta t$ to invoke the central limit theorem.
Let $\Delta t = N \delta t$ with a large integer $N$ then the sum of $N$ random spatial translations and boosts has
asymptotically a normal distribution with mean values and variances proportional to $N$ and hence to $\Delta t$.
The mean values of  ${\mathbf a}$ and  ${\mathbf u}$ are assumed to vanish. This leads to the modified
Postulate
\setcounter{post}{0}
\begin{post}\label{P2}(Final form)\\
Between the points in time $t$ and $t+\Delta t$, the total  time evolution,
in addition to the pure time evolution $U(\Delta t)$, is given by random, independent spatial translations ${\mathbf a}$
and Galilean boosts ${\mathbf u}$ which are normally distributed with zero mean values $\langle {\mathbf a}\rangle ={\mathbf 0}$
and  $\langle {\mathbf u}\rangle ={\mathbf 0}$ and variances
\begin{eqnarray}
\label{vara}
  \sigma^2({\mathbf a}) &=& \alpha\, \Delta t, \\
  \label{varu}
   \sigma^2({\mathbf u}) &=& \beta\, \Delta t
   \;,
\end{eqnarray}
where $\alpha>0$ and $\beta>0$ are free parameters of the theory $QT^\ast$.
\end{post}

In order to calculate the effect $W\mapsto W'={\mathcal T}_S W$ of the spatial quantum fluctuations ${\mathbf a}$ on the statistical operator
$W$ we assume its velocity representation
by means of the integral kernel $W({\mathbf  v},{\mathbf w})$. According to Postulate \ref{P2} and (\ref{irrepGalredv}) we obtain
\begin{eqnarray}
\nonumber
 W({\mathbf v},{\mathbf w})& \mapsto & W'({\mathbf v},{\mathbf w})=(2\pi \alpha \Delta t)^{-3/2}
 \int{\exp\left(-\frac{\sf i}{\hbar} m ({\mathbf v}-{\mathbf w})\cdot {\mathbf a}\right) }\\
 \label{effect1}
 &&
 \exp\left(-\frac{{\mathbf a}^2}{2\alpha \Delta t} \right) W({\mathbf v},{\mathbf w})\,d^3{\mathbf a} \\
 \label{effect2}
  &=& \exp\left(-\frac{m^2 \alpha \Delta t}{2 \hbar^2} ({\mathbf v}-{\mathbf w})^2\right)\,W({\mathbf v},{\mathbf w})
  \;.
\end{eqnarray}
Note that the transformation  $W\mapsto W'={\mathcal T}_S W$ preserves positivity and trace, and hence maps statistical
operators onto statistical operators, although it is not unitarily induced. Being a mixture of
unitarily induced transformations it is completely positive \cite{K83} and, moreover,  increases the von Neumann entropy
$S(W)= -k_B \mbox{Tr}(W\,\log W)$
since $W\mapsto S(W)$ is a concave function.

It will be convenient to introduce new velocity coordinates by
\begin{equation}\label{newlambdamu}
 {\boldsymbol \lambda}=\frac{1}{\sqrt{2}}({\mathbf v}+{\mathbf w}),\quad
 {\boldsymbol \mu}=\frac{1}{\sqrt{2}}({\mathbf v}-{\mathbf w}),
\end{equation}
and to rewrite (\ref{effect2}) in the form
\begin{equation}\label{effect3}
 W( {\boldsymbol \lambda},  {\boldsymbol \mu})\mapsto
 W'( {\boldsymbol \lambda},  {\boldsymbol \mu})=
 \exp\left(-\frac{m^2 \alpha \Delta t}{\hbar^2} {\boldsymbol \mu}^2\right)\,
  W( {\boldsymbol \lambda},  {\boldsymbol \mu})
  \;.
\end{equation}

Analogously, we  calculate the effect $W'\mapsto W''={\mathcal T}_B W'$ of the quantum boost fluctuations ${\mathbf u}$ on the statistical operator
$W'$ in terms of its position representation given by the integral kernel $W'({\mathbf x},{\mathbf y})$ and using (\ref{irrepGalredx}):
\begin{eqnarray}
\nonumber
 W'({\mathbf x},{\mathbf y})& \mapsto & W''({\mathbf x},{\mathbf y})=(2\pi \beta \Delta t)^{-3/2}
 \int{\exp\left(\frac{\sf i}{\hbar} m ({\mathbf x}-{\mathbf y})\cdot {\mathbf u}\right) }\\
 \label{effect4}
 &&
 \exp\left(-\frac{{\mathbf u}^2}{2\beta \Delta t} \right) W'({\mathbf x},{\mathbf y})\,d^3{\mathbf u} \\
 \label{effect5}
  &=& \exp\left(-\frac{m^2 \beta \Delta t}{2 \hbar^2} ({\mathbf x}-{\mathbf y})^2\right)\,W'({\mathbf x},{\mathbf y})
  \;.
\end{eqnarray}
Also this result will be rewritten by introducing new position coordinates
\begin{equation}\label{newxieta}
 {\boldsymbol \xi}=\frac{1}{\sqrt{2}}({\mathbf x}+{\mathbf y}),\quad
 {\boldsymbol \eta}=\frac{1}{\sqrt{2}}({\mathbf x}-{\mathbf y}),
\end{equation}
and assumes the form
\begin{equation}\label{effect6}
 W'( {\boldsymbol \xi},  {\boldsymbol \eta})\mapsto
 W''( {\boldsymbol \xi},  {\boldsymbol \eta})=
 \exp\left(-\frac{m^2 \beta \Delta t}{\hbar^2} {\boldsymbol \eta}^2\right)\,
  W'( {\boldsymbol \xi},  {\boldsymbol \eta})
  \;.
\end{equation}
Due to the similar structure also the transformation  $W'\mapsto W''={\mathcal T}_B W'$ preserves positivity and trace,
is completely positive 
and implies an increase of von Neumann entropy.

For later purposes we will prove the following
\begin{lemma}\label{L1}
\begin{enumerate}
  \item If the integral kernel $W({\mathbf x},{\mathbf y})$ of $W$ in the positional representation
  is uniformly bounded in absolute value,
  $\left|W({\mathbf x},{\mathbf y})\right| \le {\sf w}$ for all $({\mathbf x},{\mathbf y})\in {\mathbb R}^6$,
  then the mappings ${\mathcal T}_S$ and ${\mathcal T}_B$
  will decrease this bound,
  \item The analogous statement holds for the integral kernel $W({\mathbf v},{\mathbf w})$ of $W$ in the velocity representation.
\end{enumerate}
\end{lemma}
{\bf Proof}: It will suffice to prove the first statement concerning the positional representation of the integral kernel of $W$.
For the effect of boost fluctuations it is clear that ${\mathcal T}_B$ decreases the absolute value of $\left|W({\mathbf x},{\mathbf y})\right|$
since it consists of a multiplication with a positive function $\le 1$, see (\ref{effect5}).
For the effect of spatial fluctuations ${\mathcal T}_S$ we consider
\begin{eqnarray}
\nonumber
  \left|\left({{\mathcal T}_S} W \right)({\mathbf x},{\mathbf y}) \right| &\le &
  \left(2\pi \alpha \Delta t \right)^{-3/2}
  \int \exp\left(-\frac{{\mathbf a}^2}{2\alpha \Delta t} \right) \left| W({\mathbf x}-{\mathbf a},{\mathbf y}-{\mathbf a})\right|  d^3 {\mathbf a} \\
  \label{spatialcontract1}
   &\le &  \left(2\pi \alpha \Delta t \right)^{-3/2}
  \int \exp\left(-\frac{{\mathbf a}^2}{2\alpha \Delta t} \right)\, {\sf w} \, d^3 {\mathbf a} ={\sf w}
  \;.
\end{eqnarray}
\hfill$\Box$\\

\subsection{Illustration of decoherence effects}\label{sec:IDE}

\begin{figure}[ht!]
\centering
\includegraphics*[clip=true,width=1.0\columnwidth]{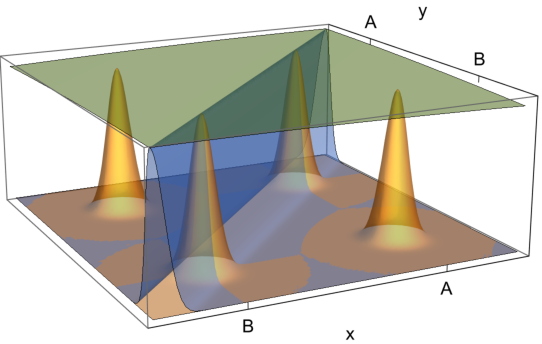}
\caption{Illustration of decoherence effects.
The integral kernel $W'(x,y)$ (dark yellow surface) corresponds to a wave function
that is strongly concentrated at the points $x=A$ and $x=B$. By multiplication
with a Gaussian damping  function that fulfills the condition $\Delta \eta \ll |A-B|$ (blue surface),
where $\Delta \eta$ is the width of the Gaussian in the direction orthogonal to the diagonal $x=y$, defined in (\ref{widthx}),
the off-diagonal peaks of $W'(x,y)$ are suppressed.
In contrast, multiplication with a Gaussian damping function that satisfies
$\Delta \eta \gg  |A-B|$ (green surface) leaves $W'(x,y)$ practically unchanged.
}
\label{FIG1}
\end{figure}

It is plausible that the damping of the ``off-diagonal" regions of the integral kernels
$W({\mathbf v},{\mathbf w})$ and $W'({\mathbf x},{\mathbf y})$ by the transformations (\ref{effect2}) and (\ref{effect5})
possibly leads to decoherence effects. Nevertheless, it will be instructive to study these effects by means of an example.

We consider only one spatial dimension $x$ and a statistical operator $W'$ with integral kernel $W'(x,y)$
given by the projector onto a wave function representing a superposition of two wave functions sharply
localized at $x=A$ and $x=B$, resp., such that $A<B$. It follows that $W'(x,y)$ almost vanishes everywhere
except for four small regions located around the points $(x=A,y=A),(x=A,y=B),(x=B,y=A)$ and $(x=B,y=B)$, see
Figure \ref{FIG1}. Through the transformation $W'\mapsto W''$, see (\ref{effect5}), the integral kernel  $W'(x,y)$ is multiplied by
the ``damping" Gaussian $\exp\left(-\frac{m^2 \beta \Delta t}{2 \hbar^2} (x-y)^2\right)$ which has the width
\begin{equation}\label{widthx}
  \Delta \eta = \frac{\hbar}{m\sqrt{\beta \Delta t}}
\end{equation}
in the $\eta$-direction orthogonal to the diagonal $x=y$.

The effect of this multiplication
depends on the size of $ \Delta \eta$ relative to $|A-B|$:
If  $\Delta \eta \gg |A-B|$ the integral kernel $W'(x,y)$ remains practically unchanged.
The converse case $\Delta \eta \ll |A-B|$ leads to a complete suppression of the off-diagonal
peaks of $W'(x,y)$ at $(x=A,y=B)$ and $(x=B,y=A)$, see Figure \ref{FIG1}. The resulting
integral kernel $W''(x,y)$ no longer represents a pure ensemble but a statistical mixture
of two pure ensembles corresponding to localized wave functions at $x=A$ or $x=B$.
In the regime between these two extremal cases $W''(x,y)$ is a mixture of ensembles
corresponding to localized wave functions or superpositions where the statistical weight
of the superpositions is gradually diminished for decreasing width  $\Delta \eta$
of the Gaussian.

Before returning to the general case, we note that a simulation of the transformation
$W'\mapsto W''$ as a single process, starting from a superposition $\phi(x)$ of
localized wave functions and performing a sequence of random boosts,
would  not change the double-peaked structure of $|\phi(x)|^2$.
There is no ``collapse of the wave packet" that accompanies the
decoherence process $W'\mapsto W''$ for the case of $\Delta \eta \ll |A-B|$.
However, we would not see this as a disadvantage of the decoherence approach but
rather as an argument against the interpretation of the
wave function as the state of an individual system, see the discussion in Section \ref{sec:S}.

\subsection{The classical world}\label{sec:CW}

In the example of the preceding subsection we have only considered the effect of random boosts
that lead to the multiplication of the integral kernel $W'({\mathbf x},{\mathbf y})$ of a statistical operator
with a Gaussian damping function $\exp[-\frac{{\boldsymbol \eta}^2}{ (\Delta \eta)^2}]$,
where ${\boldsymbol \eta}=\frac{1}{\sqrt{2}}{\mathbf x}-{\mathbf y}$.
It is clear that, analogously, random translations transform the integral kernel $W({\mathbf v},{\mathbf w})$
by a multiplication with $\exp[-\frac{{\boldsymbol \mu}^2}{ (\Delta \mu)^2}]$,
where ${\boldsymbol\lambda}$ and ${\boldsymbol\mu}$ are defined in (\ref{newlambdamu})
and
\begin{equation}\label{widthv}
  \Delta \mu = \frac{\hbar}{m\sqrt{\alpha \Delta t}}
  \;.
\end{equation}
In a similar way as sketched in subsection \ref{sec:IDE} this transformation could possibly
destroy superpositions of pure ensembles with different velocities, depending on the size of $\Delta \mu$
and the integral kernel $W({\mathbf v},{\mathbf w})$.

It seems at first sight that decoherence effects will be arbitrarily strong for increasing $\Delta t$.
However, there is a time threshold $\tau$, so that the decoherence reaches
its maximum for $\Delta t\sim\tau$ and does not increase further
when $\Delta t\gtrsim\tau$. $\tau$ can be obtained by the condition that
the two widths $\Delta \eta$ and $\Delta \mu$ satisfy the Heisenberg uncertainty relation:
\begin{equation}\label{maxtau}
  (\Delta \eta)\, (m\Delta \mu) \sim \hbar/2 \stackrel{(\ref{widthx},\ref{widthv})}{\Leftrightarrow}
  \Delta t \sim \frac{\hbar}{2m \sqrt{\alpha \beta}}=:\tau
  \;.
\end{equation}
For $\Delta t \sim \tau$, the attenuation of $W'({\mathbf x},{\mathbf y})$
near the diagonal causes a broadening of $W'({\mathbf v},{\mathbf w})$
due to the uncertainty principle, and vice versa, which brings the decoherence process to a standstill.
$\tau$ may be called the ``maximal decoherence time" due to Galilean decoherence. It is the simplest
quantity with the dimension of time that can be formed from the parameters of the quantum system and
Galilean fluctuation. In general, typical decoherence times will be shorter than $\tau$, depending of the
initial ensemble $W$.

Generalizing the example discussed in subsection \ref{sec:IDE} we expect that Galilei decoherence will
lead to a statistical operator $W$ which is approximately ``classical", i.~e.,
a mixture of pure states sharply localized in phase space (or, in this paper, position-velocity space).
An example of such pure states is provided by the projectors onto (squeezed) coherent states $|\Omega \rangle$,
see, e.~g., \cite{KS85}.
These coherent states initially will have general variances $\sigma_x^2, \sigma_v^2$ w.~r.~t.~position and velocity,
that sharply satisfy the uncertainty relation, i.~e., $(\sigma_x)\,(m\sigma_v)=\hbar/2$.
To fix the values of $\sigma_x$ and $\sigma_v$ depending on the masses $m$, we define
\begin{equation}\label{defsigma}
 \sigma_x^2:=\alpha \tau= \sqrt{\frac{\alpha}{\beta}}\frac{\hbar}{2 m},\quad
 \sigma_u^2:=\beta \tau= \sqrt{\frac{\beta}{\alpha}}\frac{\hbar}{2 m},
\end{equation}
which entails
\begin{equation}\label{sigmauncert}
   (\sigma_x)\, (m \sigma_u)=\frac{\hbar}{2}
\;.
\end{equation}

In the \ref{sec:A1} we will present some calculations which confirm the expectation that Galilean decoherence
leads to an approximately classical state if the following condition for decoherence is satisfied:
\begin{equation}\label{conddec}
{\sf M}\gtrsim \frac{1}{2}
\;.
\end{equation}
Here ${\sf M}>0$ a dimensionless parameter defined by
\begin{equation}\label{defMsf}
 {\sf M}:=\frac{\sigma_x}{\Delta \eta} \stackrel{(\ref{widthx},\ref{defsigma})}{=}
 \sqrt{\frac{m}{2\hbar}} \left(\alpha\beta\right)^{1/4}
 \stackrel{(\ref{widthv},\ref{defsigma})}{=}\frac{\sigma_v}{\Delta \mu}
\;.
\end{equation}
Note that for $\Delta t=\tau$ this parameter assumes the value
\begin{equation}\label{limitMV}
 {\sf M}\left.\right|_{\Delta t=\tau}=\frac{1}{2}
 \;.
\end{equation}

There is an extensive literature on the problem of whether quantum mechanics
contains classical mechanics as a limiting case, see \cite{S99}, Chapter X,
and in which sense decoherence plays a role in this, see \cite{S23} for a recent account.
We can only address a few aspects of this debate here, in particular the controversy
between Ballentine and Schlosshauer on the latter question \cite{WB05}, \cite{S08}, \cite{B08},
insofar as it indirectly affects issues of our paper.

First, it should be noted that in the so-called ``Ehrenfest regime" there are
approximate correspondences between classical trajectories and certain solutions
of the Schr\"odinger equation, independent of any decoherence effects.
The conditions for this are that the initial states are products of wave packets
that are sharply localized with respect to position and momentum,
and that the potential energy function is practically constant
within the widths of the wave functions. Furthermore, this approximate correspondence
is only valid for limited times, which can be relatively short for chaotic movements.
For example, the trajectory of a billiard ball after $11$ collisions is no longer uniquely
determined even macroscopically due to Heisenberg's uncertainty principle for the
initial conditions, see \cite{R67}.

In this respect, Ballentine has a point here, even if his research relates less to the
Ehrenfest regime than to the so-called ``Liouville regime" \cite{B04}.
On the other hand, representatives of decoherence theories emphasize the
aspect that an approximate reduction of classical mechanics to quantum mechanics must also
explain why certain superpositions of macroscopic states (such as states of living and dead cats)
cannot be prepared. A representative of Ballentine's position could reply to this
that such a superposition can
indeed be prepared, but in the sense of the ensemble interpretation and not in the sense of individual systems in a superposition.
We have already criticized this view in the section \ref{sec:EIAB}
and would therefore recognize the need to explain the limited
preparability of states in the CSM, but would point out that our variant of
decoherence also provides the desired explanation.

In the case of the chaotic dynamics mentioned above, the theory $QT^\ast$
leads to a transformation of the relatively rapidly spreading pure state
into a ``proper" mixture. It thus also contains the limiting case of stochastic classical mechanics,
which, as far as I can see, is not possible in the usual decoherence theories.

\section{Quantum measurement}\label{sec:QM}

A quantum measurement consists of coupling a microsystem to a macrosystem (``measuring apparatus")
such that the time evolution of the total system leads to final states that can be classically distinguished.
The problem is that, using only the pure time evolution, the final state will be an entangled superposition of
different product states. In this section we will examine the extent
to which the picture changes when Galilean fluctuations are taken into account.

We assume, for the sake of simplicity, that the Hilbert space ${\mathcal H}_S$ of the microsystem
will be finite-dimensional and that the action of the Galilei group onto ${\mathcal H}_S$ is trivial
(represented by the identity). The Hilbert space of the measuring apparatus is assumed to be of the form
\begin{equation}\label{Hilbertmeas}
{\mathcal H}_A = {\mathcal H}_R \otimes {\mathcal H}_m
\;,
\end{equation}
where ${\mathcal H}_m$ and the corresponding action of the reduced Galilei group has been defined in (\ref{Hilbert3}) and
(\ref{irrepGalredx}). ${\mathcal H}_m$ physically represents the center of mass degrees of freedom of a part of the measuring
apparatus (``pointer") that is used for the ultimate reading of the measurement result. The Hilbert space ${\mathcal H}_R$
represents the other parts of the measuring apparatus and the remaining degrees of freedom of the pointer.

The total Hilbert space is hence
\begin{equation}\label{Hilberttotal}
{\mathcal H}= {\mathcal H}_S \otimes {\mathcal H}_R \otimes {\mathcal H}_m
\;,
\end{equation}
and the pure time evolution between $t$ and $t+\Delta t$  is given by a unitary operator
$U(\Delta t): {\mathcal H} \rightarrow {\mathcal H}$. We assume that the total system is initially
in a pure product state $\Phi_0= \phi_0\otimes \Psi_0\otimes \psi_0$.
The next assumption is crucial for the further discussion and can be seen as an implicit definition
of when a macrosystem with an interaction that leads to $U(\Delta t)$ is considered a ``measuring device".
\begin{ass}\label{AM}
Using the preceding notation we assume:
 \begin{equation}\label{EAM}
  U(\Delta t)\,\left(\phi_0\otimes \Psi_0\otimes \psi_0\right) \approx {\sf e}^{{\sf i}g(\Delta t)}
  \sum_{\mu=1}^{r} c_\mu \phi_\mu\otimes \Psi_\mu\otimes \Omega_\mu
  =:\Phi(\Delta t)
  \;,
 \end{equation}
 where $c_\mu >0$ for $1\le\mu\le r$ and the coherent states $\Omega_\mu$ are macroscopically distinct, i.~e.,
 $\langle \Omega_\mu \left| \right. \Omega_\nu\rangle\approx \delta_{\mu\nu}$ for $1\le\mu,\nu\le r$
 and $g(\Delta t)$ is a real function.
\end{ass}
This assumption on the state after the pure time evolution $U(\Delta t)$ resembles the bi-orthogonal Schmidt decomposition of
a twofold product state, see \cite{P93}, Chapter 5.3, but note that in general there does not exist a Schmidt decomposition
of triple product states. Physically, the Assumption \ref{AM} means that, besides the fundamental
entanglement between the microsystem and the measuring apparatus which is represented in the superposition (\ref{EAM}),
the further entanglement between the pointer and the remaining degrees of freedom of the measuring apparatus
can be neglected.
We will not specify the form of the unitary representation of the Galilei group in the Hilbert space
${\mathcal H}_R$ but only assume, as always, that the pure time evolution $  U(\Delta t)$ commutes with the total representation
of the Galilei group restricted to translations and boosts.

The projector onto the state vector $\Phi(\Delta t)$ is hence of the form
\begin{equation}\label{WDeltat}
  W(\Delta t)= | \Phi(\Delta t)\rangle \langle \Phi(\Delta t)| =
  \sum_{\mu,\nu=1}^{r}c_\mu c_\nu | \phi_\mu\rangle \langle \phi_\nu| \otimes
  | \Psi_\mu\rangle \langle \Psi_\nu| \otimes
  | \Omega_\mu \rangle \langle \Omega_\nu|
  \;.
\end{equation}
The effect of the Galilei fluctuations will only be calculated for the last factor
$| \Omega_\mu \rangle \langle \Omega_\nu| $ of (\ref{WDeltat}).
This already shows that after the time $\Delta t$ the pure state decays into a
mixture of macroscopically distinguishable pointer states.
It seems highly plausible that additional decoherence effects cannot affect this result.

We set
\begin{equation}\label{defWss}
 W''_{\mu\nu}({\mathbf x},{\mathbf y}):= \langle {\mathbf x}\left|{\mathcal T}_B {\mathcal T}_S
 \left(| \Omega_\mu \rangle \langle \Omega_\nu|\right)
 \right| {\mathbf y}\rangle
 \;,
\end{equation}
$W''$ can be explicitly calculated with the result
\begin{eqnarray}
\nonumber
 && W''_{\mu\nu}({\mathbf x},{\mathbf y})= \left(2\pi (\alpha \Delta t+\sigma_x^2) \right)^{-3/2} \exp\Big[ \\
 \nonumber
  && -\frac{1}{8  (\alpha \Delta t+\sigma_x^2)}\left(
  2({\mathbf x}-{\mathbf x}_\mu)^2 + 2({\mathbf y}-{\mathbf x}_\nu)^2
  +\left({\mathbf x}-{\mathbf x}_\mu+{\mathbf x}_\nu-{\mathbf y}\right)^2\frac{\alpha \Delta t}{\sigma_x^2}
  \right)\\
  \nonumber
  && -\frac{\tau\,M^2}{2(\Delta t +\tau)\sigma_u^2}\left({\mathbf v}_\mu - {\mathbf v}_\nu\right)^2
  -\frac{M^2}{\sigma_x^2}\left({\mathbf x} - {\mathbf y}\right)^2
  -\frac{4{\sf i}m}{\hbar}\Big(\alpha \Delta t\left(
  {\mathbf v}_\mu\cdot\left({\mathbf x}+ {\mathbf x}_\mu+{\mathbf x}_\nu+{\mathbf y}\right)\right.  \\
  \nonumber
  &&+
 \left( {\mathbf v}_\nu\cdot\left({\mathbf x}- {\mathbf x}_\mu-{\mathbf x}_\nu-{\mathbf y}\right)
  \right)
  +2\sigma_x^2\left({\mathbf v}_\mu\cdot{\mathbf x}-{\mathbf v}_\nu\cdot{\mathbf y} \right)
  \Big)\Big]\\
  \label{Wss}
  &&
\end{eqnarray}
Here we we have set $\left|\Omega_\mu\right\rangle =\left|{\mathbf x}_\mu,{\mathbf v}_\mu\right\rangle $
for $1\le \mu\le r$.

By assumption, for $\mu \neq \nu$ the pointer states $\Omega_\mu, \Omega_\nu$ will be  macroscopically distinguishable.
This means that ${\mathbf x}_\mu,{\mathbf x}_\nu$ or ${\mathbf v}_\mu,{\mathbf v}_\nu$ will be  macroscopically distinguishable.
The latter can be translated into the condition $\left|{\mathbf v}_\mu-{\mathbf v}_\nu\right| \gg \sigma_u$, and, analogously,
$\left|{\mathbf x}_\mu-{\mathbf x}_\nu\right| \gg \sigma_x$ for the first condition.

First,
the factor $\exp\left[-\frac{\tau\,M^2}{2(\Delta t +\tau)\sigma_u^2}\left({\mathbf v}_\mu - {\mathbf v}_\nu\right)^2 \right]$ in (\ref{Wss})
implies that for $\mu\neq \nu$ the absolute values $\left|  W''_{\mu\nu}({\mathbf x},{\mathbf y})\right|$ can be,
for all practical purposes (fapp), considered as zero since $\frac{\tau\,M^2}{2(\Delta t +\tau)}\gtrsim 1$, except
for the case ${\mathbf v}_\mu\approx{\mathbf v}_\nu$. In this case we can, however, assume that
$\left|{\mathbf x}_\mu-{\mathbf x}_\nu\right| \gg \sigma_x$ and argue as follows.

We collect the Gaussian terms in (\ref{Wss}) which depend on ${\mathbf x}$ or ${\mathbf y}$ and transform
the argument of the exponential them according to

\begin{eqnarray}
\nonumber
  E_{\mu\nu} &:=& -\frac{1}{4 (\alpha\Delta t +\sigma_x^2)}
  \Big[\left({\mathbf x}-{\mathbf x}_\mu \right)^2+\left({\mathbf y}-{\mathbf x}_\nu \right)^2 \\
  \label{Emunu1}
  &&+\left({\mathbf x}-{\mathbf x}_\mu+{\mathbf x}_\nu-{\mathbf y}\right)^2\frac{\alpha \Delta t}{2\sigma_x^2} \Big]
  -\frac{M^2}{\sigma_x^2}(\left({\mathbf x}-{\mathbf x}_\mu \right)^2\\
  \nonumber
  &=&-\frac{2  M^2}{\left(8 M^2+1\right)\sigma_x^2}{\boldsymbol\eta}_{\mu\nu} ^2
  -\frac{8 M^2+1} {4 \sigma_x^2}\left({\boldsymbol\eta}
 -\frac{{\boldsymbol\eta}_{\mu\nu}}{8  M^2+1}\right)^2\\
  \label{Emunu2}
 &&-\frac{1}{4 (\alpha  \Delta t+\sigma_x^2)}({\boldsymbol \xi} -{\boldsymbol \xi}_{\mu\nu} )^2
 \;,
\end{eqnarray}
where we have used (\ref{newxieta}) and the analogous definitions
\begin{equation}\label{newxietamunu}
 {\boldsymbol \xi}_{\mu\nu}=\frac{1}{\sqrt{2}}({\mathbf x}_{\mu}+{\mathbf x}_{\nu}),\quad
 {\boldsymbol \eta}_{\mu\nu}=\frac{1}{\sqrt{2}}({\mathbf x}_\mu-{\mathbf x}_\nu)
 \;.
\end{equation}
Analogously as above, the first term in (\ref{Emunu2}) will lead to a fapp vanishing of $\left|  W''_{\mu\nu}({\mathbf x},{\mathbf y})\right|$
also in the remaining case of $\left|{\mathbf x}_\mu-{\mathbf x}_\nu\right| \gg \sigma_x$.

Summarizing, the decoherence effects due to Galilean fluctuations will lead to fapp vanishing
$ W''_{\mu\nu}({\mathbf x},{\mathbf y})$ except for the diagonal elements $ W''_{\mu\mu}({\mathbf x},{\mathbf y})$.
This implies that (\ref{WDeltat}) can be  replaced by
\begin{equation}\label{WDeltatdec}
  W''(\Delta t)\approx
  \sum_{\mu=1}^{r}c_\mu^2 | \phi_\mu\rangle \langle \phi_\mu| \otimes
  | \Psi_\mu\rangle \langle \Psi_\mu| \otimes
  | \Omega_\mu \rangle \langle \Omega_\mu|
  \;.
\end{equation}
After the time $\Delta t$ the pure state $W(\Delta t)$ decays into a statistical mixture of product states
that correspond to definite pointer states $\Omega_\mu,\,\mu=1,\ldots,r$.

\section{Example: Stern-Gerlach experiment}\label{sec:ESG}

\begin{figure}[ht!]
\centering
\includegraphics*[clip=true,width=1.0\columnwidth]{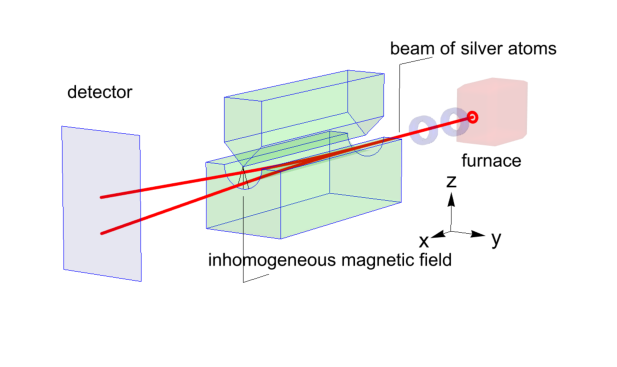}
\caption{Schematic representation of the Stern-Gerlach experiment.
}
\label{SG}
\end{figure}

\subsection{Simplified model}\label{sec:SM}

As an example to illustrate our previous considerations, we choose a simplified theoretical
description, adapted to our definitions, of the historical Stern-Gerlach experiment (1922),
in which the quantization of the electron spin was demonstrated (at least in retrospect).

A beam of silver atoms is sent through an inhomogeneous magnetic field and split into
two partial beams due to the interaction of the magnetic moment of the outer
$5s$ electron with the magnetic field, see Figure \ref{SG}.
In the simplified representation, we consider the spin of the $5s$ electron as a microsystem
with  2-dimensional Hilbert space ${\mathcal H}_S={\mathbb C}^2$.
This microsystem is initially coupled to another, auxiliary microsystem, namely the center of mass of the silver atom
with a Hilbert space ${\mathcal H}_R=L^2({\mathbb R}^3,d^3 {\mathbf r}_1)$.
The interaction with the inhomogeneous magnetic field can be described by the Pauli equation
applied to spinors in the Hilbert space ${\mathcal H}_S\otimes {\mathcal H}_R$.

After this part of the measurement the silver atom in turn interacts with a
macrosystem (``detector") and is finally detected in a rudimentary position measurement.
We model this last part of the measurement by an elastic collision of the silver atom
with another macroscopic particle $P$ (the ``pointer")
with Hilbert space ${\mathcal H}_m=L^2({\mathbb R}^3,d^3 {\mathbf r}_2)$.
$P$ is initially located at the point $(A,0,C)$ where the upwards deflected
silver atom is expected to arrive at the screen.
It must be ensured that the downwardly deflected silver atom does not interact with $P$.
Furthermore, it will be sufficient to simplify the collision as a one-dimensional problem
by considering only the coordinate in the direction of the upward deflected beam,
which we again denote by $x$
(neglecting the small $z$ component of the beam direction for this part of the calculation).

For future reference, we recall that a Gaussian wave packet satisfying the $1$-dimensional free Schr\"odinger equation
can be written in the form
\begin{eqnarray}\nonumber
  \Psi(r_i,t;m,v_i,d)&=&
  \sqrt[4]{\frac{2}{\pi }} \sqrt{\frac{d m}{2 d^2 m+{\sf i} t \hbar }}
  \exp \left(-\frac{d^2 m^2 }{4 d^4 m^2+t^2 \hbar ^2}(r_i-t v_i)^2\right)\\
  \label{Gaussian}
  &&
  \exp \left({\sf i}\frac{\left(m t r_i^2 \hbar ^2-4 d^4 m^3 v_i (t v_i-2 r_i)\right)}{8 d^4
   m^2 \hbar +2 t^2 \hbar ^3}\right)
  \;,
\end{eqnarray}
with initial value
\begin{equation}\label{Gaussian0}
 \Psi(r_i,0;m,v_i,d)=  \frac{1}{\sqrt[4]{2 \pi }}\,{\frac{1}{\sqrt{d}}} \exp\left[{-\frac{r_i^2}{4 d^2}+{\sf i}\frac{ m v_i r_i}{\hbar }}\right]
 \;.
\end{equation}
Here $r_i$ is one coordinate of the position vector ${\mathbf r}$, $m$ denotes the mass of the particle, $v_i$ a velocity parameter such that
$m v_i$ will be the expectation value of $r_i$-momentum and $d$ will be the standard deviation
of the $r_i$-position observable at $t=0$. Generally we have
\begin{equation}\label{standarddevoft}
  \sigma_{r_i}^2(t)=d^2+\frac{t^2 \hbar ^2}{4 d^2 m^2}
  \;,
\end{equation}
which describes the well-known spreading of the wave packet with increasing time.
This spreading can be neglected for short times, more precisely, for
\begin{equation}\label{dissneg}
  t \ll \frac{2 d^2 m}{\hbar}
  \;.
\end{equation}

A general element of the product space ${\mathcal H}_S\otimes {\mathcal H}_R={\mathbb C}^2\otimes L^2({\mathbb R}^3,d^3{\mathbf r}_1)$
can be written in the form of a spinor ${\Psi_\uparrow ({\mathbf r}_1,t) \choose \Psi_\downarrow ({\mathbf r}_1,t)}$.
We will first concentrate on the first component $\Psi_\uparrow ({\mathbf r}_1,t)$.
Initially, $\Psi_\uparrow ({\mathbf r}_1, t=t_1)$ will be chosen as a Gaussian wave packet at rest.
During the first part of the measurement, corresponding to its flight through the magnet,
the silver atom will be accelerated into the direction, say, of increasing $z_1$.
The exact calculation of the motion of the silver atom in an inhomogeneous magnetic field would be
difficult, but we may assume that, at the end of the first part,
at time $t_2$, the first component $\Psi_\uparrow ({\mathbf r}_1,t_2)$ is approximately given by
\begin{equation}\label{psit2}
  \Psi_\uparrow ({\mathbf r}_1,t_2)\approx  \Psi(x_1,0;m_1,u,d_1)
\, \Psi(y_1,0;m_1,0,d_1)\, \Psi(z_1,0;m_1,v,d_1)  \;,
\end{equation}
where we have set ${\mathbf r}_1=\left(x_1,y_1,z_1\right)$, inserted the mass $m_1$ of the silver atom into (\ref{Gaussian}) and
chosen our space-time coordinate system such that  $t_2=0$  and the wave packet has the position expectation
value $\langle {\mathbf r}_1\rangle=(0,0,0)$ at $t=t_2=0$. The width $d_1>0$ is left unspecified and will be chosen later.

The next part of the measurement is the free time evolution of the silver atom described by $\psi_\uparrow ({\mathbf r}_1,t)$ until
it collides with the macroscopic particle $P$ at $x_1=A$ .
At time $t=t_2=0$ the macroscopic particle with mass $m_2$
will be represented by the Gaussian wave packet written in product form
\begin{eqnarray}\nonumber
  \psi_\uparrow({\mathbf r}_2,t_2)&=& \Psi(x_2-A,0;m_2,0,d_2)\,
 \Psi(y_2,0;m_2,0,d_2)\, \Psi(z_2-C,0;m_2,0,d_2)
 ,\\
 \label{psit2}
 &&
\end{eqnarray}
where we have set ${\mathbf r}_2=\left(x_2,y_2,z_2 \right)$ and introduced $C$ as the vertical coordinate of
the point where the upwards deflected beam hits the screen.
The distance $A>0$ and the widths $d_1,\,d_2$ have to be chosen such that the two wave packets,
that of the silver atom and that of the pointer, practically do not overlap at time $t=0$.

As noted above, we will simplify the treatment of the collision by considering only the horizontal
coordinates $x_1$ of the silver atom and $x_2$ of the pointer.
Then the collision dynamics is most conveniently calculated by passing to center of mass and relative coordinates
\begin{equation}\label{cmrel}
 X:= \frac{m_1 x_1+m_2 x_2}{m_1+m_2},\quad x:=x_1-x_2
 \;,
\end{equation}
since the motion w.~r.~t.~$X$ will be free if the interaction only depends on $x$.
The inverse transformation is given by
\begin{equation}\label{cmrelinv}
 x_1= X+\frac{m_2}{m_1+m_2}x,\quad  x_2= X-\frac{m_1}{m_1+m_2}x
 \;.
\end{equation}

Here we encounter the following problem: If we insert (\ref{cmrelinv})
into $ \Psi_\uparrow (x_1,t)   \psi_\uparrow(x_2,t)$ the result can, in general, not be factored
into a wave function depending on $X$ and another one depending on $x$.
This would call our assumption \ref{AM} into question.
Fortunately, we can achieve the factorization
\begin{equation}\label{factor}
 \Psi_\uparrow (x_1,t)   \psi_\uparrow(x_2,t)= \Phi_\uparrow(X,t) \phi_\uparrow(x,t)
\end{equation}
for all $t$ by the special choice
\begin{equation}\label{specialchoice}
 d_2=d_1\,\sqrt{\frac{m_1}{m_2}}
 \;,
\end{equation}
as can be shown by a short computation.
This condition can be reformulated equivalently by saying that the ``dissipation time"
$t^{(i)}=\frac{2 d_i^2\, m_i}{\hbar},\,i=1,2,$ of the two wave packets is the same.

The Schr\"odinger equation describing the collision between the silver atom
and the pointer can then be decoupled into two equations of the form
\begin{eqnarray}
\label{SE1}
 {\sf i}{\hbar}\frac{\partial}{\partial t} \Phi_\uparrow(X,t)&=& -\frac{\hbar^2}{2M} \frac{\partial^2}{\partial X^2} \Phi_\uparrow(X,t),\\
  \label{SE2}
  {\sf i}{\hbar}\frac{\partial}{\partial t} \phi_\uparrow(x,t)&=& -\frac{\hbar^2}{2\mu} \frac{\partial^2}{\partial x^2} \phi_\uparrow(x,t)
  +V(x)\phi_\uparrow(x,t)
  \;.
\end{eqnarray}
Here $M$ and $\mu$ are the total and the reduced mass, resp.~, defined in the usual way by
\begin{equation}\label{defMnu}
  M:= m_1+m_2,\quad \mu:= \frac{m_1 m_2}{m_1+m_2}
  \;,
\end{equation}
and $V(x)$ is the interaction potential.
(\ref{SE1}) describes a free time evolution for the center of mass wave function $ \Phi_\uparrow(X,t)$.
For the interaction potential we choose a repulsive delta function of the form
\begin{equation}\label{deltapot}
  V(x) = {\sf v}_0\, \delta(x), \quad \mbox{such that } {\sf v}_0\gg \hbar v
  \;,
\end{equation}
which entails that the transmission probability can be neglected
and the silver atom will be reflected by the pointer with probability $\approx 1$.
Otherwise we  would not be able to maintain our assumption \ref{AM}.
Hence the solution of (\ref{SE2}) is practically identical to a perfect reflection at
an infinite wall potential and is thus given by a superposition of an incoming wave packet $\phi_\uparrow(x,t)$
and a reflected one $-\phi_\uparrow(-x,t)$ for $x<0$ and a vanishing wave function for $x\ge 0$.
At the time $t=t_2=0$ practically only the incoming wave is present and at the time $t\ge t_3:= 2 A/v$
only the reflected wave. This presupposes that the dispersion of the wave packet
$\phi_\uparrow(x,t)$ has not become too large at $t=t_3$ which will be checked later when
concrete number values have been chosen.

We conclude that at time $t\ge t_3$ the collision is practically finished and that the total
wave function will be of the form
\begin{equation}\label{wavet3}
 -\Phi_\uparrow(X,t) \phi_\uparrow(-x,t) =: \Psi_\uparrow^+(x_1,t)\,\psi_\uparrow^+(x_2,t)
 \;,
\end{equation}
where we have again invoked the condition (\ref{specialchoice}) in order
to factorize the result in the specified way.

In order to confirm the assumption \ref{AM} it has to be shown that,
for $t\ge t_3$,  the final pointer state $\psi_\uparrow^+(x_2,t)$
is approximately a coherent state (up to a time-dependent phase factor).
To this end we will use the condition (\ref{dissneg}) for the pointer state
which can be translated into $\theta \ll 1$ if the dimensionless parameter
$\theta$ is implicitly defined by setting
\begin{equation}\label{theta}
 t= \frac{d_2^2 m_2}{\hbar}\,\theta
 \;.
\end{equation}
The details of the calculation will be given in  \ref{sec:A2}.

So far we have only considered the first component of the spinor
${\Psi_\uparrow ({\mathbf r}_1,t) \choose \Psi_\downarrow ({\mathbf r}_1,t)}$.
It would be a straight forward task to repeat the foregoing considerations for
the second component by considering the silver atom with spin down being
deflected downwards and colliding with a second pointer.
But in order to keep the example as simple as possible we do not consider a second pointer
and only argue that in this case the silver atom practically will not interact
with the first pointer which hence is, for $\theta\ll 1$, represented by a
Gaussian wave packet $\psi_\downarrow^+({\mathbf r}_2,t)$ at rest.

It remains to confirm that, for some time $t > t_3$ the two pointer states
$\psi_\uparrow^+(x_2,t)$ and $\psi_\downarrow^+(x_2,t)$, which according to \ref{sec:A2} can be
approximated by coherent states $\Omega_\uparrow$ and $\Omega_\downarrow$,
can be macroscopically distinguished. Recall that $\Omega_\uparrow= |x_0,v_0\rangle$
with $x_0=A\frac{m_2-m_1}{m_2+m_1}+\frac{2 m_1 v}{m_1+m_2} t\approx A +\frac{2 m_1 v}{m_1+m_2} t$
for $m_1 \ll m_2$, see (\ref{x0v0}), whereas $\Omega_\downarrow= |A,0\rangle$.
Hence the condition of $\Omega_\uparrow$ and $\Omega_\downarrow$ being macroscopically distinct
is equivalent to
\begin{equation}\label{macdis}
  \frac{2 m_1 v}{m_1+m_2} t \gg d_2
  \;.
\end{equation}
The consistency of the various assumptions will be examined in the following subsection.

\subsection{Choice of concrete values}\label{sec:CV}

It is the aim of this subsection to show, by choosing concrete values for the  physical
quantities of the simplified Stern-Gerlach experiment and for the free parameters $\alpha$ and $\beta$ of the theory $QT^\ast$,
that the various assumptions we made in the previous subsection are consistent
and also in agreement with the general theory of Galilean decoherence that we have developed.

Thus let us assume that a silver atom with mass $m_1=1.79\times 10^{-25}\, kg$ moves with a horizontal
velocity of $u=600\, \frac{m}{s}$ through a magnet of length $L=0.25\, m$ with
an inhomogeneity of $\frac{\partial B}{\partial z}=120 \,\frac{T}{m}$.
The time of flight through the magnet is therefore $|t_1|=4.17\times 10^{-4} \,s$.
The spin $1/2$ electron
in the outer shell of the silver atom has a magnetic moment of $1\, \mu_B= 9.274\times 10^{-24} \,\frac{J}{T}$.
Due to the vertical acceleration in the inhomogeneous magnetic field,
the silver atom finally reaches a vertical velocity of
\begin{equation}\label{vnum}
v= \frac{\frac{\partial B}{\partial z}\mu_B |t_1|}{m_1}=2.5905\,\frac{m}{s}
\;.
\end{equation}

After leaving the magnet, the silver atom moves horizontally for the time $\frac{t_3}{2}=3.333\times 10^{-4}\,s$
until it hits the screen at a horizontal distance of $A=0.2\,m$.

Next we choose the mass of the macroscopic particle $P$ as $m_2= 10^8\, m_1 = 1.79\times 10^{-17}\,kg$,
approximately corresponding to the mass of a $18$ corona viruses.
Further let
\begin{equation}\label{alphabeta}
 \alpha=2.5\times 10^{-13}\,\frac{m^2}{s}, \quad \mbox{and  }\beta=5.0\times 10^{-17}\, \frac{m^2}{s^3}
 \;,
\end{equation}
which, by virtue of (\ref{maxtau}), (\ref{specialchoice}) and (\ref{defsigma}), implies
\begin{eqnarray}\label{d2}
 d_2&=&\sigma_x= \left({\frac{\alpha}{\beta}}\right)^{1/4}\sqrt{\frac{\hbar}{2\,m_2}}=3.618\times 10^{-8}\,m \;,\\
 \label{d1}
 d_1&=& d_2 \sqrt{\frac{m_2}{m_1}}=10^4\,d_2=3.618\times 10^{-4}\,m  \;,\\
 \label{sigmau}
 \sigma_u&=& \left({\frac{\beta}{\alpha}}\right)^{1/4}\sqrt{\frac{\hbar}{2\,m_2}}=5.116\times 10^{-10}\,\frac{m}{s}\;,\\
 \label{tau}
 \tau&=&\frac{\hbar}{2\,m_2\,\sqrt{\alpha\,\beta}}=5.235\times 10^{-3}\,s
 \;.
\end{eqnarray}

Using these values we will check the consistency of our assumptions.
First we estimate the vertical distance $\Delta z=2C$ between the two partial beams when they
hit the screen as
\begin{equation}\label{delza}
 \Delta z = 2 t_3 v = 3.454\times 10^{-3}\,m \approx 9.5 \,d_1 = 9.5\times 10^4 \,d_2
 \;,
\end{equation}
It can therefore actually be ruled out that the silver atom deflected
downwards interacts with the particle $P$, which is initially located at $z_2=\frac{1}{2} \Delta z=C$.

Next we assume that the duration $\Delta t$ of the total measuring process
will be  $\Delta t=10^2\,\tau=0.5235\,s$. The times $|t_1|$ and $t_3$
for the first parts of the measurement can therefore be neglected in comparison with $\Delta t$.
During the time $\Delta t$, the pointer moves by the horizontal distance
\begin{equation}\label{deltaz}
  \Delta x \approx \frac{2 m_1 u \Delta t}{m_1+m_2} =6.28\times 10^{-6} m \approx 174 \, d_2
  \;.
\end{equation}
This means that the two coherent states $\Omega_\uparrow$ and $\Omega_\downarrow$
can indeed be considered as macroscopically distinct and in this sense the condition (\ref{macdis})
can be regarded as fulfilled.

The condition $\theta\ll 1$ which guarantees that the wave packet of the pointer
practically will not spread during the measurement process is satisfied in the sense that
\begin{equation}\label{thetall}
  \Delta t =0.5235\,s \ll \frac{2 d_2^2\,m_2}{\hbar} =70.71\,s
  \;.
\end{equation}
With the shorter collision time $t_3=6.667\times 10^{-4}\,s$
the wave packet of the silver atom will not spread either due to (\ref{specialchoice}).

Finally, we consider the decoherence process after the time $\tau$.
The width $\Delta \eta$ in the Gaussian function (\ref{effect6}) assumes the value
\begin{equation}\label{Deltaetaex}
  \Delta \eta = \frac{\hbar}{m_2\sqrt{\beta \tau}}=7.2353\times 10^{-8}\,m =0.0115176\,\Delta x
  \;.
\end{equation}
Hence the superposition of the
two pointer states $\Omega_\uparrow$ and $\Omega_\downarrow$  is damped by a
factor which is fapp zero. In contrast, the analogous width $\Delta \eta_1$
for the damping factor of the silver atom, calculated for the whole duration $\Delta t$ of the measurement,
assumes the value
\begin{equation}\label{Deltaetae1}
  \Delta \eta_1 = \frac{\hbar}{m_1\sqrt{\beta \Delta t}}=0.72353\,m =2,000\,d_1
  \;,
\end{equation}
which means that Galilean decoherence of the state of the silver atom can be safely neglected.

To summarize, we can say that a consistent choice of physical quantities
is possible for the Stern-Gerlach experiment, which confirms our
theory of decoherence by Galilean fluctuations.

\section{Discussion and Summary}\label{sec:S}

There are numerous versions of the measurement problem in quantum mechanics
in terms of its formulation and possible solutions. What justifies the publication of yet another version?
On the one hand, it is the impression that the actual debate is too one-sidedly
based on an ``individual system interpretation" of quantum states.
Furthermore, the discussion mostly seems to assume that macrosystems
can also be described by quantum mechanics without further modifications,
a position that we referred to in the Introduction as ``Q-monism".

We have therefore presented and discussed the ensemble interpretation in
relative detail in the section \ref{sec:EI}, following Ludwig's approach.
However, in our opinion, the thesis of Ballentine, another representative of the
ensemble interpretation, that the measurement problem is a pseudo-problem,
is not tenable, see subsection \ref{sec:EIAB}.

There is a connection between the two points, ensemble interpretation and Q-monism.
If, following Ludwig's approach, one formulates an axiomatic basis of quantum mechanics
with the basic concepts of ``preparation" and ``registration procedures",
there are obvious problems in applying these basic concepts to the preparation and measurement
of all macroscopic quantum states. In this case, Ludwig therefore speaks of an ``extrapolation"
of quantum mechanics \cite{L87}. However, one does not necessarily have to follow
this ``skeptical semantics", since there are other examples of physical theories with successful extrapolation:
Nobody will doubt that the diameter of the sun is $1,392,700 \mbox{km}$
only because terrestrial scales would evaporate inside the sun.

The focus of our work is on the sketch of a modified quantum theory $QT^\ast$,
in which the measurement problem can be solved if the ensemble interpretation is adopted.
The modification consists in the postulate of a global quantum fluctuation
with respect to the position and velocity of quantum systems,
which is effective in addition to the pure time evolution, whereby an arrow of time is distinguished.
The average effect of these fluctuations leads to decoherence effects
dependent on the mass $m$ of the system with a (maximal) decoherence time $\tau(m)$.
This does not allow an absolute distinction between micro- and macrosystems,
but a distinction that depends on whether $\tau(m)\gg \Delta t$ or $\tau(m)\ll \Delta t$
applies with a time $\Delta t$ typical for the duration of experiments.
The restriction of $QT^\ast$ to macrosystems leads to a classical statistical mechanics (CSM)
with a phase space with Cartesian $({\mathbf p},{\mathbf q})$-coordinates.
This is at least highly plausible, even if we have not worked out the details.
The approximate reduction $QT^\ast \rightarrow CSM$ is not surprising,
but essentially a consequence of the postulate of phase space fluctuations,
analogous to the identification of a ``pointer basis" by the form of the
interaction in decoherence theories.
In our case, the role of the pointer basis is taken over by the
overcomplete basis of coherent phase space states $|\Omega\rangle$.
The analogous restriction to microsystems leads to the usual quantum theory $QT$,
since Galilean decoherence can be neglected for microsystems and times $\Delta t\ll \tau(m)$.
As the time evolution of $QT^\ast$ differs from that of $QR$, these theories will not
be empirically equivalent. Especially for systems at the boundary between micro and macro
decoherence effects should be measurable and provide estimates for the parameters $\alpha$ and $\beta$.

The quantum mechanical measurement process was treated in the theory $QT^\ast$
under the restrictive condition that the entangled total state
after the pure measurement interaction has the form of a superposition of triple product states.
This can be heuristically justified by the fact that the transition to a mixture of
macroscopically distinguishable pure states by virtue of
Galilean decoherence is then particularly easy to handle.
To show the consistency of the various assumptions with each other and
with the choice of the free parameters in $QT^\ast$,
we have considered the spin measurement in a Stern-Gerlach magnet as an example.
The tricky point here is the interaction of the deflected silver atom with a
macroscopic detector, which we have roughly modeled as an elastic collision with a macroscopic particle.

We should emphasize that the theory $QT^\ast$ in no way describes the
occurrence of a concrete measurement result in a single experiment.
This would not be expected simply because of the underlying ensemble interpretation of quantum states.
Rather, the Galilean decoherence transforms the entangled state
into a ``proper" mixture, which allows an ignorance interpretation.
This ignorance is eliminated by reading the measurement result.

Finally, we would like to discuss the role of the $QT^\ast$ theory.
It is initially intended as a toy theory that can be used to
demonstrate the fundamental possibility of a theory comprising $QM$ and $CSM$.
On the other hand, there are also physical arguments that make a
 modified quantum theory along the lines of $QT^\ast$ does not appear to be completely arbitrary.
We recall that in Ludwig's statistical interpretation of quantum mechanics,
the effect of Galilean transformations on states (or, alternatively,
on effects) was indirectly defined by the effect of these transformations
on (classical) coordinate frames. The construction manuals for
preparation apparatus refer to these coordinate frames.
In this respect, the arbitrary accuracy with which a Galilean
transformation and its representation is specified is an extrapolation
of a classical inaccuracy. However, this accuracy can be questioned in terms of
semantic consistency (this time in a different context than mentioned in the Introduction):
The devices, scales, clocks and optical measuring instruments,
for the numerical determination of a Galilean transformation are
again quantum objects. They therefore exhibit typical uncertainties
and fluctuations that stand in the way of an arbitrarily precise determination of a Galilean transformation.

We mention as an additional source of uncertainty the classical description
of (non-relativistic) gravity by a Newton-Cartan geometry with a curved connection \cite{H64},
which is reminiscent of Penrose's idea. This replaces the definition of
Galilean transformations in the classical sense with path-dependent parallel transport of Galilean frames.

However, a mathematical formulation of this fundamental uncertainty of
Galilean transformations is completely open. The requirement of
Galilean fluctuations between two points in time $t$ and $t+\delta t$
can be seen as a first attempt at such a formulation.
One can therefore speculate whether other formulations
lead to similar decoherence effects and thus explain the emergence of a
classical world for times larger than  $\tau(m)$.

\section*{Acknowledgment}
I would like to thank Thomas Br\"ocker for numerous discussions on the topics covered in this paper and for his encouragement in writing it.

\appendix

\section{Decay into a mixture of coherent states}\label{sec:A1}

We consider the (squeezed) coherent states $|\Omega\rangle=\left|{\mathbf x}_0,{\mathbf v}_0\right\rangle$ defined in the position representation by
\begin{equation}\label{defcoh1}
\langle {\mathbf x} |\Omega\rangle =\langle {\mathbf x}|{\mathbf x}_0,{\mathbf v}_0\rangle
=\left(2\pi \sigma_x^2\right)^{-3/4}
 \exp\left[-\frac{({\mathbf x}-{\mathbf x}_0)^2}{4 \sigma_x^2}\right]
\exp\left[\frac{\sf i}{\hbar} m {\mathbf x}\cdot {\mathbf v}_0 \right]
\;.
\end{equation}
This definition is in accordance with \cite{KS85} except that we consider the position-velocity space
instead of phase space. Due to the completeness relation
\begin{equation}\label{complete}
 \Eins=\int |\Omega\rangle \langle \Omega|\, d\Omega
\end{equation}
we may represent any statistical operator $W$ in the form
\begin{equation}\label{repW}
 W=\int\int |\Omega'\rangle \langle \Omega'| W | \Omega\rangle \langle \Omega| \,d\Omega\, d\Omega'
 \;.
\end{equation}
The corresponding mixture of coherent states has the form
\begin{equation}\label{mixtW}
 \widetilde{W}=\int|\Omega\rangle \langle \Omega| W | \Omega\rangle \langle \Omega| \,d\Omega\
  \;,
 \end{equation}
which is formally obtained from  (\ref{repW}) by setting
\begin{equation}\label{Wapprox}
  \langle \Omega'| W | \Omega\rangle =\langle \Omega| W | \Omega\rangle \,\delta(\Omega',\Omega)
  \;.
\end{equation}
Hence an ensemble $W$ is approximately a mixture of coherent states, $W\approx \widetilde{W}$, if
the integral kernel $ \langle \Omega'| W | \Omega\rangle $ is sharply peaked about the diagonal $\Omega'=\Omega$.

The following calculations will show that this is the case for the statistical operator $W''$
resulting after the two transformations $W\mapsto W'={\mathcal T}_S\,W$
and $W'\mapsto W''={\mathcal T}_B\,W'$ considered in Section \ref{sec:PR}.
The treatment is not mathematically rigorous; if only for the reason that the formulation
``$\langle \Omega'| W | \Omega\rangle$ is  sharply peaked about  the diagonal $\Omega'=\Omega$"
should be mathematically made more precise.
Moreover, will restrict ourselves to the case where the absolute values of the
integral kernels $W({\mathbf x},{\mathbf y})$ and $W({\mathbf v},{\mathbf w})$ are uniformly bounded.
This would be certainly true for examples using superpositions of Gaussians multiplied
with exponential phase factors, but not generally for
square-integrable $W({\mathbf x},{\mathbf y})$.
As a side note, the rigorous use of integral kernels is tricky; even the innocent-looking equation
$\mbox{Tr } W = \int W({\mathbf x},{\mathbf x})\,d^3{\mathbf x}$
requires some non-trivial mathematical interpretation work, see \cite{B88}.
\\

First we consider
$\langle \Omega'| W''|\Omega\rangle=\langle {\mathbf x}'_0,{\mathbf v}'_0| W''|{\mathbf x}_0,{\mathbf v}_0\rangle$:
\begin{eqnarray}
\label{Wss1}
  \langle \Omega'| W''|\Omega\rangle
 &=&
  \int\int \langle \Omega'| {\mathbf x}
  \rangle W''({\mathbf x},{\mathbf y}) \langle {\mathbf y}|\Omega\rangle d^3{\mathbf x}\,d^3{\mathbf y} \\
  \nonumber
   &=& \left(2\pi \sigma_x^2\right)^{-3/2}\int\int \exp\left[-\frac{({\mathbf y}-{\mathbf x}_0)^2}{4\sigma_x^2}
   -\frac{({\mathbf x}-{\mathbf x}'_0)^2}{4\sigma_x^2} \right]\\
   && \exp\left[\frac{\sf i}{\hbar}m\left({\mathbf y}\cdot{\mathbf v}_0-{\mathbf x}\cdot{\mathbf v}'_0 \right) \right]
   \,W''({\mathbf x},{\mathbf y})\,d^3{\mathbf x}\,d^3{\mathbf y}\\
   \label{Wss2}
   \nonumber
   &=&\left(2\pi \sigma_x^2\right)^{-3/2}\int\int \exp\left[-\frac{({\boldsymbol \xi}-{\boldsymbol \xi}_0)^2}{4\sigma_x^2}
   -\frac{({\boldsymbol \eta}-{\boldsymbol \eta}_0)^2}{4\sigma_x^2} \right]\\
   \label{Wss3}
   && \exp\left[\frac{\sf i}{\hbar}m\left({\boldsymbol \xi}\cdot{\boldsymbol \mu}_0-{\boldsymbol \eta}\cdot{\boldsymbol \lambda}_0 \right) \right]
   \,W''({\boldsymbol \xi},{\boldsymbol \eta})\,d^3{\boldsymbol \xi}\,d^3{\boldsymbol \eta}
   \;.
\end{eqnarray}

Here we have, additionally to (\ref{newxieta}) and (\ref{newlambdamu}), defined
\begin{equation}\label{newxieta0}
 {\boldsymbol \xi}_0=\frac{1}{\sqrt{2}}({\mathbf x}_0+{\mathbf x}'_0),\quad
 {\boldsymbol\eta}_0=\frac{1}{\sqrt{2}}({\mathbf x_0}-{\mathbf x}'_0)
 \;,
\end{equation}
and
\begin{equation}\label{newlambdamu0}
 {\boldsymbol \lambda}_0=\frac{1}{\sqrt{2}}({\mathbf v}_0+{\mathbf v}'_0),\quad
 {\boldsymbol \mu}_0=\frac{1}{\sqrt{2}}({\mathbf v}_0-{\mathbf v}'_0)
 \;.
\end{equation}
Using
\begin{eqnarray}\label{WssWs1}
 W''({\boldsymbol \xi},{\boldsymbol \eta})
& \stackrel{(\ref{effect6})}{=}&
 \exp\left(-\frac{m^2 \beta \Delta t}{\hbar^2} {\boldsymbol \eta}^2\right)\,
  W'( {\boldsymbol \xi},  {\boldsymbol \eta})\\
 \label{WssWs2}
 & \stackrel{(\ref{defMsf})}{=}&
  \exp\left(-{\sf M}^2\frac{{\boldsymbol \eta}^2}{\sigma_x^2}\right)\,
  W'( {\boldsymbol \xi},  {\boldsymbol \eta})
\end{eqnarray}
we collect all terms of (\ref{Wss3}) arising in the $\eta$-integral:
\begin{eqnarray}\nonumber
I_\eta &:=&
\left(2\pi \sigma_x^2\right)^{-3/2}
 \int \exp\left[-\frac{({\boldsymbol \eta}-{\boldsymbol \eta}_0)^2}{4\sigma_x^2}
  -\frac{\sf i}{\hbar}m{\boldsymbol \eta}\cdot{\boldsymbol \lambda}_0 \right]\\
  \label{etaint}
  &&
  \exp\left(-{\sf M}^2\frac{{\boldsymbol \eta}^2}{\sigma_x^2}\right)\,
  W'( {\boldsymbol \xi},  {\boldsymbol \eta})\,d^3{\boldsymbol\eta}
  \end{eqnarray}
According to our assumptions and Lemma \ref{L1} there exists a least upper bound  ${\sf w}>0$ such that
$\left|   W'( {\boldsymbol \xi},  {\boldsymbol \eta})\right|
\le \left|   W( {\boldsymbol \xi},  {\boldsymbol \eta})\right|\le {\sf w}$
for all $( {\boldsymbol \xi},  {\boldsymbol \eta})\in {\mathbb R}^6$.
This enables the following estimate:
\begin{eqnarray}
\label{est1}
  \left|I_\eta \right|
  &\le&
  \left(2\pi \sigma_x^2\right)^{-3/2}
 \int \exp\left[-\frac{({\boldsymbol \eta}-{\boldsymbol \eta}_0)^2}{4\sigma_x^2}
 -{\sf M}^2\frac{{\boldsymbol \eta}^2}{\sigma_x^2}\right]\,
 {\sf w}\,d^3{\boldsymbol\eta}
 \\
 \nonumber
 &=&
 \left(2\pi \sigma_x^2\right)^{-3/2}
  \exp\left[-\frac{{\sf M}^2}{(1+4 {\sf M}^2)\sigma_x^2}{\boldsymbol \eta}_0^2 \right]\, {\sf w}\,
  \\
 \label{est2}
 && \int \exp\left[ -\frac{\frac{1}{4}+{\sf M}^2}{\sigma_x^2}\left({\boldsymbol \eta}-{\boldsymbol \eta}_0 \right)^2 \right]\,
d^3{\boldsymbol\eta}\\
 \label{est3}
&=&\frac{1}{8 \pi ^{3/2} \left(4 {\sf M}^2+1\right)^{3/2}} \exp\left[-\frac{{\sf M}^2}{(1+4 {\sf M}^2)\sigma_x^2}{\boldsymbol \eta}_0^2 \right]\, {\sf w}
\;.
 \end{eqnarray}

In the next step we insert this result into the remaining $\xi$-integration and obtain
\begin{eqnarray}\nonumber
 \left| \langle \Omega'| W''|\Omega\rangle \right|
 &\le &
 \frac{1}{8 \pi ^{3/2} \left(4 {\sf M}^2+1\right)^{3/2}} \exp\left[-\frac{{\sf M}^2}{(1+4 {\sf M}^2)\sigma_x^2}{\boldsymbol \eta}_0^2 \right]\, {\sf w}
 \\
  \label{xint1}
  &&
 \int
  \exp\left[-\frac{({\boldsymbol \xi}-{\boldsymbol \xi}_0)^2}{4\sigma_x^2}\right]
   \,d^3{\boldsymbol \xi}
 \\
  \label{xint2}
  &=&
   \frac{ \sigma_x^{3}}{ \left(4 {\sf M}^2+1\right)^{3/2}} \exp\left[-\frac{{\sf M}^2}{(1+4 {\sf M}^2)\sigma_x^2}{\boldsymbol \eta}_0^2 \right]\, {\sf w}
   \;.
\end{eqnarray}

If $\Delta t$ assumes the maximal decoherence time $\tau$ we have ${\sf M}^2=\frac{1}{4}$, see (\ref{limitMV}),
and for $1/4\le {\sf M}^2 < \infty$ the factor $\frac{{\sf M}^2}{(1+4 {\sf M}^2)}$ in the Gaussian (\ref{xint2})
varies from $1/8$ to $1/4$.
The radius of the tube around the diagonal ${\mathbf x}_0={\mathbf x}'_0$ where
$ \left| \langle \Omega'| W''|\Omega\rangle \right|$ cannot be neglected is hence of the order of $\sigma_x$.
This shows that $\langle \Omega'| W''|\Omega\rangle$ is sharply concentrated at ${\boldsymbol \eta}_0={\mathbf 0}$, or,
by (\ref{newxieta0}) equivalently, at ${\mathbf x}_0={\mathbf x}'_0$.
It is remarkable that the mentioned radius of the tube is small but not arbitrarily small.
This is the first part of what we want to show.

The remaining task is to show that $\langle \Omega'| W''|\Omega\rangle$ is sharply concentrated at ${\boldsymbol \mu}_0={\mathbf 0}$.
Here we can argue quite analogously to the previous discussion
by using the velocity representation instead of the position representation
and the fact that the order of action of ${\mathcal T}_S$ and ${\mathcal T}_B$ can be swapped to obtain $W''$.
Hence the radius of the tube around the diagonal ${\mathbf v}_0={\mathbf v}'_0$ where
$ \left| \langle \Omega'| W''|\Omega\rangle \right|$ cannot be neglected is of the order of $\sigma_u$.

Summarizing, $\langle \Omega'| W'' | \Omega\rangle$ is  sharply peaked about  the diagonal $\Omega'=\Omega$
and due to the Galilean fluctuations the statistical operator can be approximated by a mixture of classical states.
We have shown that this happens for $\Delta t=\tau$ but for concrete examples the effective decoherence time
may be much smaller than the maximal decoherence time $\tau$,

\section{Approximation by coherent states in the Stern-Gerlach example}\label{sec:A2}

We explicitly calculate the pointer state $\psi_\uparrow^+(x_2,t)$
after the collision with the silver atom
and find that it is of the form
\begin{equation}\label{psiupex}
\psi_\uparrow^+(x_2,t)=f(t) \exp\left[R(x_2,t)+{\sf i}\, I(x_2,t) \right]
\;,
\end{equation}
where $f(t)$ is a complex function used for normalization (except for a time-dependent phase factor)
and $R(x_2,t)$ and $I(x_2,t)$ are real functions of the following form:
\begin{eqnarray}\label{Rex1}
  R(x_2,t)&=&-\frac{d_2^2 m_2^2}{4 d_2^4 m_2^2+t^2 \hbar^2}\left(x_2-A\frac{m_2-m_1}{m_2+m_1}-\frac{2 m_1 v}{m_1+m_2} t\right)^2\\
\nonumber
  &=& \left( -\frac{1}{4 d_2^2} +\frac{\theta^2}{16 d_2^2} +O(\theta^4)   \right)
  \left(x_2-A\frac{m_2-m_1}{m_2+m_1}-\frac{2 m_1 v}{m_1+m_2} t\right)^2\;,\\
    \label{Rex2}
  &&
\end{eqnarray}
where $theta$ was the dimensionless parameter defined in (\ref{theta}).
Moreover,
\begin{equation}\label{Im}
  I(x_2,t)=I_1(t) x_2 + I_2(t) x_2^2
  \;,
\end{equation}
where
\begin{eqnarray}
\label{Im1a}
 I_1(t)&=& \frac{8 d_2^4 m_1 m_2^3 v+A (m_1-m_2)m_2 t \hbar^2}{(m_1+m_2)\hbar(4 d_2^4 m_2^2 +t^2 \hbar^2)} \\
 \label{Im1b}
  &=& \frac{2 m_1 m_2 v}{(m_1+m_2)\hbar} + \frac{A(m_1-m_2)}{4d_2^2(m_1+m_2)}\theta +O\left(\theta^2\right)
  \;,
\end{eqnarray}
and
\begin{equation}
\label{Im2}
 I_2(t)=\frac{m_2 t \hbar}{8 d_2^4 m_2^2+ 2 t^2 \hbar^2}
  = \frac{\theta}{8 d_2^2}+O\left(\theta^3\right)
 \;.
\end{equation}
We conclude that, if all terms of linear or higher order in $\theta$ are neglected, the wave packet
$\psi_\uparrow^+(x_2,t)$ can be, up to a time-dependent phase factor,
approximated by a coherent state $\Omega_\uparrow=|x_0,v_0\rangle$ with
\begin{equation}\label{x0v0}
x_0=A\frac{m_2-m_1}{m_2+m_1}+\frac{2 m_1 v}{m_1+m_2} t,\quad \mbox{and  } v_0=\frac{2 m_1 v}{m_1+m_2}
\;.
\end{equation}
It has to be checked that the standard deviation $\sigma_x=d_2$ of this coherent state
is in accordance with the other data of the Stern-Gerlach example
and the general theory outlined in this paper, see subsection \ref{sec:CV}.

\section*{References}


\begin{thebibliography}{15}


\bibitem{W85}
C.~F.~von Weizs\"acker,
\textit{Aufbau der Physik}, Carl Hanser, M\"unchen, Wien, 1985.

\bibitem{N32}
J.~von Neumann,
\textit{Matnematische Grundlagen der Quantenmechanik}, Springer, Berlin, 1932,
(English translation: ``Mathematical foundations of quantum
mechanics", Princeton University Press, Princeton, New Jersey, 1955).

\bibitem{L54}
G.~Ludwig,
\textit{Die Grundlagen der Quantenmechanik},
Springer, Berlin, 1954.

\bibitem{L87}
G.~Ludwig,
\textit{An Axiomatic Basis for
Quantum Mechanics,
Volume 2,
Quantum Mechanics and Macrosystems},
Springer, New York, 1987.

\bibitem{LB39}
F.~London, E.~Bauer, and P.~Langevin,
\textit{La th\'{e}orie de 1'observation en m\'{e}canique quantique},
Hermann, Paris, 1939.

\bibitem{E57}
H.~Everett III,
``Relative state" formulation of quantum mechanics,
\textit{Rev. Mod. Phys.} {\bf 29} (3), 454 -- 462, (1957)

\bibitem{dW68}
B.~S.~deWitt,
 The Everett-Wheeler interpretation of quantum mechanics,
 in:
C.~M.~deWitt and J.~A.~Wheeler (eds.)
\textit{Battelle Rencontres, 1967 lectures in mathematics and physics},
 Benjamin, New York, 1968, pp.~318 -- 332.

\bibitem{BB66}
D.~Bohm and J.~Bub,
A proposed solution of the measurement problem in quantum mechanics by a hidden variable theory,
\textit{Rev. Mod. Phys.} {\bf 38} (3), 453 -- 469, (1966)


\bibitem{G84}
R.~B.~Griffiths,
Consistent Histories and the Interpretation of Quantum Mechanics,
\textit{J. Stat. Phys.} {\bf 36} (1-2), 219 -- 272, (1984)

\bibitem{GRW86}
G.~C.~Ghirardi, A.~Rimini, and T.~Weber,
Unified dynamics for microscopic and macroscopic systems,
\textit{Phys. Rev. D} {\bf 34} (2), 470 -- 491, (1986)


\bibitem{Getal96}
D.~Giulini, E.~Joos, C.~Kiefer, J.~Kupsch,
I.-O.~Stamatescu, and H.~D.~Zeh (eds.),
\textit{Decoherence
and the Appearance
of a Classical World
in Quantum Theory},
Springer, Berlin Heidelberg New York, 1996.

\bibitem{P89}
R.~Penrose,
\textit{The Emperor's New Mind: Concerning Computers, Minds and The Laws of Physics},
Oxford University Press, Oxford (UK), 1989.





\bibitem{B70}
L.~E.~Ballentine,
The statistical interpretation of quantum mechanics,
\textit{Rev. Mod. Phys.} {\bf 42} (4), 358 -- 381, (1970)

\bibitem{B15}
L.~E.~Ballentine,
\textit{ Quantum Mechanics: A Modern Development}, 2nd edition,
World Scientific, Singapore, 2015.


\bibitem{L85}
G.~Ludwig,
\textit{An Axiomatic Basis for
Quantum Mechanics,
Volume 1,
Derivation of the Hilbert Space Structure},
Springer, New York, 1985.


\bibitem{HW92}
D.~Home and M.~A.~B.~Whitaker,
Ensemble interpretations of quantum mechanics. A modern perspective,
\textit{Phys. Rep.} {\bf 210} (4), 223 -- 317, (1992)


\bibitem{P93}
A.~Peres,
\textit{Quantum Theory: Concepts and Methods}, Kluwer, Dordrecht, 1993.


\bibitem{KS67}
S.~Kochen and E.~P.~Specker,
The Problem of Hidden Variables in Quantum Mechanics,
\textit{J. Math. Mech.} {\bf 17} (1), 59 -- 87, (1967)


\bibitem{L76}
J.-M.~L\'{e}vy-Leblond,
Nonrelativistic Particles and Wave Equations,
\textit{Commun. math. Phys.} {\bf 6}, 286 -- 311, (1967)


\bibitem{K83}
K.~Kraus,
\textit{States, Effects, and Operations Fundamental Notions of Quantum Theory}, 
Lecture Notes in Physics 190,  Springer, Berlin, 1983.


\bibitem{KS85}
J.~R.~Klauder and B.~S.~Skagerstam,
\textit{Coherent states: applications in physics and mathematical physics}, World scientific, Singapore, 1985.


\bibitem{S99}
E.~Scheibe,
\textit{Die Reduktion physikalischer Theorien, Teil II: Inkommensurabilit\"at und Grenzfallreduktion}, Springer, Berlin, 1999.


\bibitem{S23}
P.~Strasberg,
Classicality with(out) decoherence: Concepts, relation to
Markovianity, and a random matrix theory approach,
\textit{ SciPost Phys.} {\bf 15} (1), 024, (2023)


\bibitem{WB05}
N.~Wiebe and L.~E.~Ballentine,
Quantum mechanics of Hyperion,
\textit{Phys. Rev. A} {\bf 72},  022109, (2005)


\bibitem{S08}
M.~Schlosshauer,
Classicality, the Ensemble Interpretation,
and Decoherence: Resolving the Hyperion Dispute,
\textit{Found. Phys.} {\bf 38}, 796 -- 803, (2008)


\bibitem{B08}
L.~E.~Ballentine,
Classicality without Decoherence:
A Reply to Schlosshauer,
\textit{Found. Phys.} {\bf 38}, 916 -- 922, (2008)

\bibitem{R67}
D.~J.~Raymond,
How Determinate is the ``Billiard Ball Universe"?,
\textit{Am. J. Phys.} {\bf 35} (2), 102 -- 103, (1967)


\bibitem{B04}
L.~E.~Ballentine,
Quantum-to-classical limit in a Hamiltonian system,
\textit{Phys. Rev. A} {\bf 70} (3), 032111, (2004)


\bibitem{B88}
C.~Brislawn,
Kernels of trace class operators,
\textit{Proc. Am. Math. Soc.} {\bf 104} (4), 1181 -- 1190, (1988)

\bibitem{H64}
P.~Havas,
Four-dimensional formulations of Newtonian mechanics and their relation to the special and general theory of relativity,
\textit{Rev. Mod. Phys.} {\bf 36} (4), 938 -- 965, (1964)


\end{thebibliography}
\end{document}